\documentclass[aps,prfluids,preprint,groupedaddress]{revtex4-2}

\usepackage[x11names]{xcolor}
\usepackage{graphicx}
\usepackage{amsmath, amssymb}
\usepackage{mathtools}
\usepackage{enumerate}
\usepackage[colorlinks]{hyperref}
\usepackage{bm}

\begin{document}

\newcommand{\enstrophy}{\mathcal{Z}}
\newcommand{\dissls}{\ell_\nu} 
\newcommand{\power}{\mathcal{P}_\mathcal{F}}
\newcommand{\rfric}{\alpha_\nu^{\Upsilon}}
\newcommand{\turnover}{t_L}
\newcommand{\dteval}{\Delta t_{\mbox{eval}}}
\newcommand{\Teval}{T_{\mbox{eval}}}
\newcommand{\bcdot}{\bm{\cdot}}

\title{Online learning of subgrid-scale models for quasi-geostrophic turbulence in planetary interiors}


\author{Hugo Frezat}
\email[]{hugo.frezat@gmail.com}
\author{Thomas Gastine}
\author{Alexandre Fournier}
\affiliation{Université Paris Cité, Institut de physique du globe de Paris, CNRS, F-75005 Paris, France}


\date{\today}

\begin{abstract}
Machine learning approaches to subgrid-scale (SGS) modeling are now well established in atmospheric and oceanic applications. 
Among these, online end-to-end learning,  where the differentiable solver participates in the training, has shown particular promise. 
Yet existing studies are largely restricted to idealised periodic domains, with no mechanical boundaries, precluding them from addressing the dynamics of bounded rotating flows relevant to planetary interiors. 
Here we consider two-dimensional quasi-geostrophic turbulence in a rapidly rotating annular bounded domain.
We examine three configurations varying the geometry of the container and the rotation rate. 
The system exhibits key features such as zonal jets, Rossby waves, and, in the spherical shell geometry, a slow quasi-periodic inward drift of the jets. The spectral properties of the zonal and non-zonal flow can be understood in the framework of zonostrophic turbulence theory. We develop a differentiable solver for this system, which allows us to train SGS models online, over a time span of one turnover time, using coarse-grained data from direct numerical simulations. 
In all cases, a SGS model trained on a single turnover time accurately reproduces global integrated diagnostics, energy spectra as well as long-term dynamical behaviours, such as jet migration, occurring on timescales which exceed the training window by one order of magnitude.
The online-trained model further outperforms classical hyperdiffusivity and Leith closure schemes, for which reducing the radial resolution is impractical. The resulting speed-up paves the way to further investigations of long-term geophysical fluid processes beyond the reach of direct numerical simulations. 

\end{abstract}


\maketitle

\section{Introduction \label{sec:introduction}}
The interior of active planets and stars hosts turbulent flow operating over a wide range of time and length scales. In the Earth's fluid outer core for example, the convective flow of liquid iron sustains the Earth's magnetic field against ohmic decay. This process, referred to as the geodynamo, has been at work for over 3.5 billion years. Measurements of the present and past variability of the geomagnetic field point to characteristic time scales ranging from a few years to millions of years \citep[see the reviews][and references therein]{landeau2022sustaining,finlay2023gyres}.
Likewise, the flow in the outer core is expected to span a vast range of length scales; the thickness of viscous boundary layers is not expected to exceed 1 m, while core-flow studies have evidenced a planetary-scale eccentric gyre whose size is commensurate with that of the core \citep{pais2008quasi}. Observation of the geodynamo is limited and our understanding of its behavior has benefited from numerical simulations, following the pioneering work of \citet{glatzmaier1995three}. Standing open questions remain, however, such as the origin of geomagnetic polarity reversals and the reasons explaining the variability of occurrence of these extreme events over Earth's evolution. Addressing these questions using numerical models is hampered by the range of time and length scales necessary to achieve this task, as performing direct numerical simulations of the geodynamo with meter-scale resolution over geological time scales is unfeasible. In practice, numerical studies of reversals are performed at intermediate resolution \citep[e.g.][]{wicht2016gaussian}; they favor long-term, large-scale dynamics over fast, small-scale dynamics, under the tacit assumption that small-scale flow does not impact the long-term, large-scale geomagnetic variability.

In order to deal with the unresolved, subgrid-scale (SGS) flow in geodynamo simulations, traditional approaches have relied on eddy-diffusivity models and hyperdiffusivity schemes. The simplest eddy-diffusivity models prescribe enhanced scalar diffusivities, while more sophisticated approaches such as the Smagorinsky model \citep{smagorinsky1963general} determine diffusivity from invariants of the strain rate tensor. Hyperdiffusivity is commonly applied along the periodic directions of a domain, where spectral methods naturally lend themselves to enhanced dissipation at small scales \citep{maltrud1991energy,chen2005large}. For three-dimensional spherical dynamo studies based on spherical harmonic expansions, hyperdiffusivity is typically implemented by increasing diffusivities above some threshold harmonic degree \citep{glatzmaier1995three,kuang1999numerical,nataf2015turbulence,aubert2017spherical}. However, these models are inherently anisotropic: diffusion along the bounded radial direction remains unchanged, and the scheme cannot account for non-local spectral interactions.A more physically-motivated alternative, the similarity model \citep{germano1986proposal,leonard1975energy,bardina1980improved}, appeared promising for capturing anisotropic turbulence in dynamo settings. Studies in magneto-convection under rapid rotation \citep{buffett2003comparison} and planar dynamo models \citep{matsui2005sub} showed that similarity models could reproduce the anisotropy of SGS fluxes better than scalar approaches. Subsequent work introducing commutation error corrections and dynamic closure \citep{matsui2007commutation,matsui2007dynamic,matsui2012large} confirmed their relevance in spherical shell geometry. Nevertheless, these models have not been implemented in state-of-the-art spherical dynamo codes.

An alternative option, based on machine learning (ML) using DNS data), has recently come to the fore. This trending topic is reviewed by \citet{brunton2020machine,vinuesa2022enhancing}. These ML components have shown their relevance in several applications, including atmospheric circulation \citep{yuval2020stable,arcomano2023hybrid}, oceanic flows \citep{bolton2019applications} or sea-ice dynamics \citep{finn2023deep}, but to our knowledge, applications to flows in planetary interiors remain to be seen. Beyond traditional SGS terms coming from coarse-graining, data-driven approaches have also been proposed to learn parametrizations for more complex or less directly observable subgrid processes. In particular, ML models have been trained to emulate the effects of cloud formation, precipitation, and radiative transfer \citep[e.g.][]{rasp2018deep,dueben2018challenges,yuval2021use}.
In these works, the ML component has been trained ``offline'', by learning the SGS terms independently of their interaction with the solver dynamics. When these learned components are plugged back into a simulation, the lack of coupling with the governing equations during training can potentially cause instabilities and accumulation of small-scale energy. Using differentiable programming \citet{gelbrecht2023differentiable} allowed models to be trained end-to-end, also referred to as ``online learning'' \citep{rasp2020coupled}. This new paradigm has been applied incrementally to more complex physical systems, including two-dimensional turbulence \citep{kochkov2021machine,frezat2022posteriori,yan2025adjoint} and, more recently, to full weather and climate forecasting on the sphere \citep{kochkov2024neural}.
Note, however, that ``online''-trained models that target SGS fluxes from a high-resolution reference simulation have only been applied to idealized domains without boundaries, thereby avoiding commutation errors that may arise inside or in the vicinity of the boundary layers.

Our goal in this paper is to assess the ML potential to describe SGS fluxes in fluid planetary interiors. To that end, we consider a prototype consisting of the QG equations in a container with rigid boundaries. Flow is driven by a network of vorticity pumps, which makes it possible to control the scale of energy injection \citep{lemasquerier2023zonal}.
This two-dimensional (2D) model retains several key dynamical features relevant to rapidly rotating geophysical flows. In particular, it supports the spontaneous formation of zonal jets ---large-scale, quasi-steady east–west flows arising from the inverse energy cascade and the anisotropy imposed by the variation of the height of the container with the distance to the rotation axis, the so-called $\beta$-effect--- as well as the combination of Rossby waves and turbulence, which govern the propagation of vorticity anomalies and ensure momentum transfer across the domain. The rigid top and bottom boundaries introduce Ekman boundary layers, whose frictional torques control the saturation amplitude of the large-scale flows. In addition, spatial variations of $\beta$ may yield long-term temporal variability of the zonal jets, such as the quasi-periodic inward jet migration reported by \citet{rotvig2007multiple} in spherical geometry. Poleward migrating jets have also been observed in numerical simulations of Earth's atmospheric jets \citep[e.g.][]{chemke2015poleward}. Although still a matter of debate \citep{lonner2022planetary}, one candidate mechanism behind this drift could stem from an asymmetry in the eddy momentum flux convergence between the inward and the outward flank of a zonal jet \citep{chemke2015poleward,cope2021dynamics}.
The interplay between these mechanisms ---zonal jet formation, wave propagation, Ekman friction, and jet migration--- makes this model an idealized, yet non-trivial, test bed for SGS parametrization in the context of rapidly rotating planetary flows. This model also provides an interpretative framework relevant to the experiments described in \cite{cabanes2017laboratory,lemasquerier2023zonal} and \cite{zhu2025zonal}.

In this study we develop a differentiable spectral solver of the QG equations and use it to train, in an end-to-end online fashion, neural-network-based predictions of subgrid-scale fluxes that arise from spatial coarse-graining. Section \ref{sec:numerical-model} presents the physical system and its spectral discretization.
In Sec. \ref{sec:implicit-subgrid-scale-learning} we introduce a Galerkin projection for generating coarse-grained data that are preserving the
boundary conditions, and expose the neural-network architecture. 
We next evaluate in Sec. \ref{sec:results} the trained models across three experiments spanning distinct dynamical regimes, in terms of integrated statistics, energy spectra, Rossby-wave dynamics, and the long-term behaviour of zonal jets. 
Results and perspectives are finally discussed in
Sec. \ref{sec:discussion}.

\section{The physical model and its numerical approximation \label{sec:numerical-model}}

\subsection{Governing equations \label{sec:governing_equations}}
\begin{figure}
    \begin{minipage}{.45\textwidth}
        (\textit{a}) \\
        \centerline{\includegraphics[width=\linewidth]{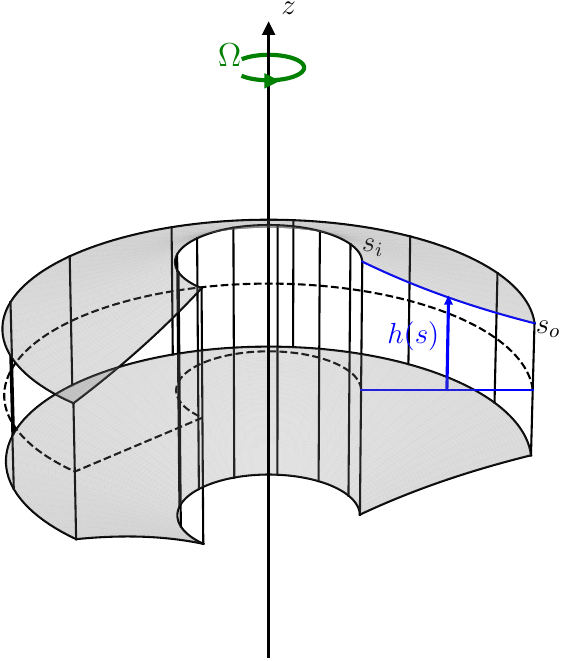}}
    \end{minipage}
    \hfill 
    \begin{minipage}{.45\textwidth}
        (\textit{b}) \\
        \centerline{\includegraphics[width=\linewidth]{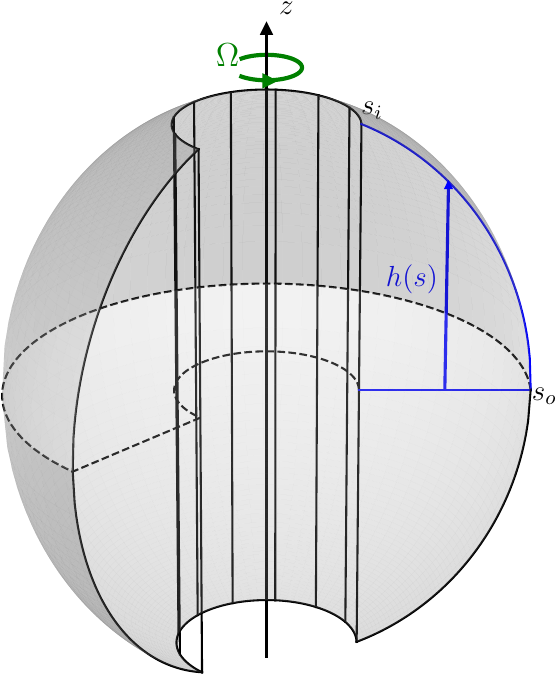}}
    \end{minipage}
    \caption{(\textit{a}) Exponential and (\textit{b}) spherical container geometries. The shape is prescribed through the half-height $h(s)$, where the cylindrical radius $s$ varies between $s_i$ and $s_o$. The container rotates at constant angular velocity $\Omega$ about the $z$ axis. \label{fig:container-shape}}
\end{figure}

We study the dynamics of an incompressible fluid of constant kinematic viscosity $\nu$ enclosed in a container rotating at constant angular velocity $\Omega$. The direction of rotation defines the $z$-axis; the container is assumed to be symmetric about this axis, and about a plane orthogonal to $z$, which we shall refer to as the equatorial plane. In the following, we operate in cylindrical coordinates $(s,\varphi,z)$ and consider the quasi-geostrophic (QG) approximation of the Navier--Stokes equations, which is expected to be valid in the limit of rapid rotation \citep[e.g.][]{aubert2003quasigeostrophic,guervilly2017multiple,lemasquerier2023zonal}.
This approximation implies that the flow is invariant along the direction of rotation and that its complete characterization can be performed in the equatorial domain, defined by the intersection of the volume of the container and the equatorial plane; consequently, field variables depend only upon $s$ and $\varphi$. In this study, the equatorial domain is an annulus of inner radius $s_{i}$ and outer radius $s_{o}$ with a radius ratio $s_{i} / s_{o} = 0.35$. The shape of the container is further specified by the half-height of the container $h$, which is a function of the sole cylindrical radius $s$. In the following we shall consider two shapes: the exponential container and the spherical shell; see Fig.~\ref{fig:container-shape}(\textit{a}) and Fig.~\ref{fig:container-shape}(\textit{b}), respectively.
In this paper, we adopt a dimensionless formulation of the QG equations using the annulus gap $d = s_{o} - s_{i}$ as the reference length scale and the viscous dissipation time $d^2 / \nu$ as the reference time scale.
Consequently, the non-dimensional Ekman number describing the ratio of viscous forces to Coriolis forces is given by $E = \nu / \Omega d^{2}$. The continuity equation under the QG approximation \citep{gillet2006quasi} then reads
\begin{equation}
    \frac{1}{s} \frac{\partial(s u_{s})}{\partial s} + \frac{1}{s} \frac{\partial u_{\varphi}}{\partial \varphi} + \beta u_{s} = 0
\end{equation}
where $u_{s}$ and $u_{\varphi}$ are the radial and azimuthal velocities, respectively. The three-dimensional geometry of the fluid container is reflected by the $\beta$ parameter which depends on the half-height $h$ at the cylindrical radius $s$,
\begin{equation}
    \beta = \frac{1}{h} \frac{\mathrm{d} h}{\mathrm{d} s}.
\end{equation} 
The two-dimensional nature of the problem makes it convenient to adopt a vorticity-streamfunction formulation, as described by \citet{aubert2003quasigeostrophic,gillet2006quasi} and implemented by \citet{gastine2019pizza} in the reference open-source code \texttt{pizza} \footnote{T. Gastine, \url{https://github.com/magic-sph/pizza}}. The cylindrically-radial and azimuthal velocity components are hence expressed using a streamfunction $\psi$ such that 
\begin{align}
    u_{s} &= \frac{1}{s} \frac{\partial \psi}{\partial \varphi}, \label{eq:us-from-psi} \\
    u_{\varphi} &= \langle u_{\varphi} \rangle_{\varphi} - \frac{\partial \psi}{\partial s} - \beta \psi. \label{eq:uphi-from-psi}
\end{align}
The azimuthal velocity $u_\varphi$ is explicitly decomposed into a mean zonal flow denoted by the angle brackets $\langle \cdots \rangle_\varphi$ and a non-zonal part, in order to ensure the periodicity of pressure \citep{plaut2002low}. The axial vorticity $\omega$ is then expressed by
\begin{equation}
    \omega = \frac{1}{s} \frac{\partial (s \langle u_{\varphi} \rangle_{\varphi})}{\partial s} - \frac{1}{s} \frac{\partial (\beta s \psi)}{\partial s} - \nabla^{2} \psi. \label{eq:omega-from-psi}
\end{equation}
Fluid motions are driven by a prescribed vorticity source term $\mathcal{F}$, such that the equations for the axial vorticity and the mean zonal velocity under the QG approximation read
\begin{align}
    \frac{\partial \omega}{\partial t} + \bm{\nabla} \bcdot (\bm{u} \omega) &= \frac{2}{E} \beta u_{s} + \mathcal{F} + \nabla^{2} \omega - \Upsilon \omega \label{eq:sph-qg-omega},\\
    \frac{\partial \langle u_\varphi \rangle_{\varphi}}{\partial t} + \langle u_s \omega \rangle_{\varphi} &= -\frac{\langle u_\varphi \rangle_{\varphi}}{s^2} + \nabla^{2} \langle u_\varphi \rangle_{\varphi} - \Upsilon \langle u_\varphi \rangle_{\varphi},
    \label{eq:sph-qg-uphi-axisymmetric}
\end{align}
supplemented with appropriate boundary and initial conditions to be defined below.
\begin{table}
    \begin{center}
    \begin{tabular}{l c c c}
    Geometry & $h$ & $\beta$ & $\Upsilon$ \\
    \hline
    Exponential  & $\exp(\beta s)$ & $\beta$ & $\dfrac{(1 + \beta^{2}h^{2})^{1/4}}{E^{1/2}h}$ \\
     & ($\beta \in \mathbb{R}^{-}$) & &  \\ 
    Spherical & $(s_{o}^{2} - s^{2})^{1/2}$ & $-\dfrac{s}{h^{2}}$ & $\dfrac{s_{o}^{1/2}}{E^{1/2}h^{3/2}}$ \\
    \end{tabular}
    \caption{Analytical expression of geometry-dependent quantities; half-height $h$, $\beta$ and Ekman pumping contribution $\Upsilon$ for exponential and spherical containers.}
    \label{tab:geometric-quantities}
    \end{center}
\end{table}
The analytical expressions for half-height $h$, $\beta$ and Ekman pumping $\Upsilon$ depend on the geometry of the 3D container (here, exponential or spherical) and are given in Table~\ref{tab:geometric-quantities}. 
The contribution due to the Ekman pumping follows from the generic expression given by \citet{greenspan1968theory}, while its derivation for the exponential container is provided in Appendix~\ref{app:ekman-exponential}; the spherical container case being detailed in Appendix~A of \citet{schaeffer2005quasigeostrophic}. In both cases, the complete expression contains more than the term proportional to $\omega$ retained in this study, as it also includes terms linear in $u_s$ and $u_\varphi$. In preliminary calculations carried out with the reference \texttt{pizza} code \citep{gastine2019pizza}, we found those to have a negligible impact on the statistical properties of the flow compared to the term in $\omega$. Consequently, we decided for computational convenience to retain that latter term only.

\begin{figure}
    \centering
    \includegraphics[width=0.45\linewidth]{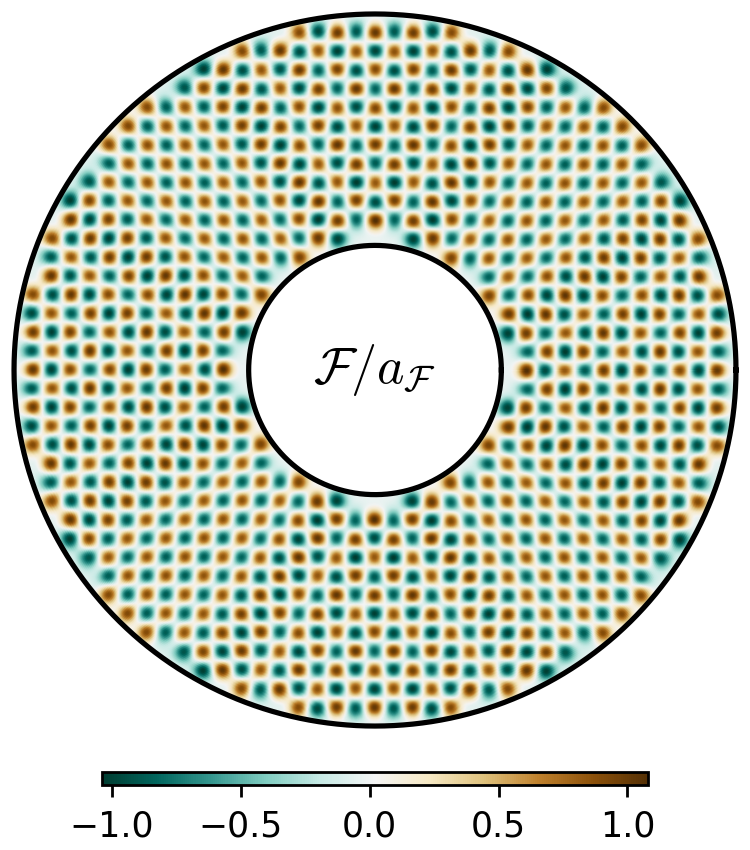}
    \caption{Cartesian forcing pattern $\mathcal{F}$ (\ref{eq:forcing}) as described by \citet{lemasquerier2023zonal} producing inlet-outlet vorticity sources. Vortex spacing $\Delta_{\mathcal{F}} = 0.08$, radius $\ell_{\mathcal{F}} = 0.04$ and amplitude $a_{\mathcal{F}} = 2 \times 10^{10}$ are the same  for all configurations studied. The magnitude of the pumps on this figure is normalised between $1$ and $-1$ for cyclonic and anticyclonic vortices, respectively. \label{fig:qg-forcing}}
\end{figure}

The stationary source term $\mathcal{F}$ consists of $N$ Gaussian sources of positive or negative vorticity, of characteristic half-width $\ell_{\mathcal{F}}$, that mimics the experimental setup of \citet{lemasquerier2023zonal},
\begin{equation}
    \mathcal{F}(x, y) = a_{\mathcal{F}} \sum_{i = 1}^{N} (-1)^{i} \exp \left[ - \left(\frac{x - x_{i}}{\ell_{\mathcal{F}}} \right)^{2} - \left(\frac{y - y_{i}}{\ell_{\mathcal{F}}} \right)^{2} \right],
    \label{eq:forcing}
\end{equation}
where $(x_i, y_i)$ are the non-dimensional Cartesian coordinates of the center of each vortex,  and $a_{\mathcal{F}}$ is the forcing amplitude. Throughout this study, we keep the same forcing structure and amplitude. The number of vortices $N$ is controlled by the vortex spacing $\Delta_{\mathcal{F}} = 0.08$, while each individual vortex half-width $\ell_{\mathcal{F}}$ is set to $0.04$. Figure~\ref{fig:qg-forcing} shows the corresponding distribution of vorticity sources over the annular domain.

Finally, we assume that the fluid is initially at rest and we prescribe no-slip mechanical boundary conditions at both boundaries, such that
\begin{equation}
    \psi = \frac{\partial \psi}{\partial s} = \langle u_{\varphi} \rangle_{\varphi} = 0 \, \text{at} \, s = s_{i},s_{o}.
    \label{eq:boundary-conditions}
\end{equation}

\subsection{Numerical discretization}
The set of equations \eqref{eq:sph-qg-omega}--\eqref{eq:boundary-conditions} is discretized using a pseudo-spectral Fourier--Chebyshev method, following the implementation of the open-source code \texttt{pizza} \citep{gastine2019pizza}.
In the azimuthal direction, fields are expanded onto a truncated real Fourier basis of $N_{m}$ modes evaluated at $N_{\varphi} = 3N_{m}$ equally-spaced grid points, so that the quadratic nonlinear terms are dealiased.
In the radial direction, a Chebyshev collocation method is employed on a grid of $N_{s}$ Gauss--Lobatto points, whose non-uniform spacing --- with clustering near both radial boundaries --- is a key feature of the discretization and will be relevant when discussing the coarse-graining procedure in Sec. \ref{subsec:coarse-graining}.
Substituting the Fourier--Chebyshev expansions into \eqref{eq:sph-qg-omega}--\eqref{eq:sph-qg-uphi-axisymmetric} yields, for each azimuthal wavenumber $m > 0$, a dense $(2N_{s} \times 2N_{s})$ linear system coupling the Chebyshev coefficients of the vorticity $\hat{\omega}_{mn}$ and streamfunction $\hat{\psi}_{mn}$ and an additional $(N_{s} \times N_{s})$ system for the mean zonal velocity, each of these enforcing boundary conditions \eqref{eq:boundary-conditions} via the tau method.

The resulting semi-discrete system is advanced in time with the third-order BPR353 implicit-explicit Runge--Kutta (IMEX-RK) scheme of \citet{boscarino2013implicit}, selected on the basis of the comparative assessment of \citet{gopinath2022assessment}.
As detailed in Sec. \ref{sec:implicit-subgrid-scale-learning}, any subgrid-scale correction $\mathcal{M}$ is naturally accommodated as an additional explicit source term, requiring three function evaluations per time step within the BPR353 stages.

Full details of the Fourier--Chebyshev discretization, the collocation matrices, and temporal discretization are provided in Appendix~\ref{app:numerical-discretisation}.

\section{Implicit subgrid-scale learning \label{sec:implicit-subgrid-scale-learning}}
To fully resolve the spatial scales in a certain configuration regime, we must run spectral simulations at DNS resolution, which in practice requires at least $N_m$ truncated Fourier modes and $N_s$ Chebyshev coefficients. We want to run ``reduced'' simulations that maintain this accuracy for coarser resolutions, i.e., $\overline{N_{m}} < N_m$ and/or $\overline{N_{s}} < N_{s}$. From now on, we employ overbars to denote the unknown fields at coarser resolution. We know that without any correction, the reduced simulations will likely accumulate energy at the smallest resolved scale, or possibly lead to a numerical blow-up in the worst case scenario. To account for the missing energy transfers, corrections terms, or subgrid-scale (SGS) terms $\tau$ must be included in the reduced (or filtered) equations,
\begin{align}
    \frac{\partial \overline{\omega}}{\partial t} + \bm{\nabla} \bcdot (\overline{\bm{u}}\,\overline{\omega}) &= \frac{2}{E} \beta \overline{u_{s}} + \overline{\mathcal{F}} + \nabla^{2} \overline{\omega} - \Upsilon \overline{\omega} + \tau_{\omega} \label{eq:sph-qg-omega-coarse}, \\
    \frac{\partial \langle \overline{u_\varphi} \rangle_{\varphi}}{\partial t} + \langle \overline{u_s}\,\overline{\omega} \rangle_{\varphi} &= -\frac{\langle \overline{u_\varphi} \rangle_{\varphi}}{s^2} + \nabla^{2} \langle \overline{u_\varphi} \rangle_{\varphi} - \Upsilon \langle \overline{u_\varphi} \rangle_{\varphi} + \tau_{\langle u_{\varphi} \rangle_{\varphi}} \label{eq:sph-qg-uphi-axisymmetric-coarse}.
\end{align}
By construction, at the resolution of the DNS, the correction $\tau = [\tau_{\omega}, \tau_{\langle u_{\varphi} \rangle_{\varphi}}]$ vanishes and \eqref{eq:sph-qg-omega-coarse}--\eqref{eq:sph-qg-uphi-axisymmetric-coarse} is equivalent to the original set of Eqs. \eqref{eq:sph-qg-omega}--\eqref{eq:sph-qg-uphi-axisymmetric}.  
The purpose of this work is to use ML to learn a model $\mathcal{M}$ that produces an approximation of the SGS terms $\tau$, given the knowledge of the reduced-resolution variables $\overline{\mathbf{z}}$,
\begin{equation}
    \mathcal{M}\left(\overline{\mathbf{z}}\right) \approx \tau.
\end{equation}

\subsection{Coarse-graining in Fourier--Chebyshev space \label{subsec:coarse-graining}}
\begin{figure}
    \centering
    \includegraphics[width=0.6\linewidth]{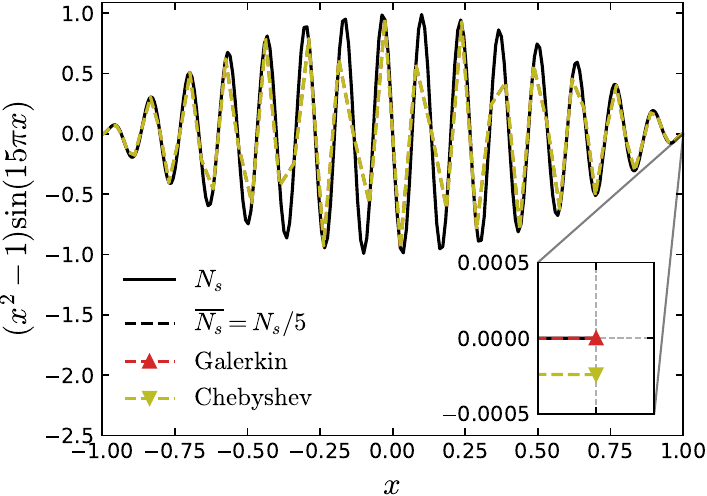}
    \caption{Example of the coarse-graining process for an analytical function $z(x) = (x^{2} - 1) \sin (15 \pi x)$ with $N_{s} = 300$ and $\overline{N_{s}} = 60$ using Chebyshev coefficients and using the boundary-preserving Dirichlet--Neumann Galerkin basis. The zoom inset clearly illustrates that the Chebyshev truncation does not preserve the boundary conditions $z(\pm 1) = 0$. \label{fig:coarse-graining}}
\end{figure}

Now, in order to generate a dataset to set-up the learning problem, we must design a composite transform $\mathcal{T}$ that maps direct-resolution fields $f$ to reduced-resolution fields $\overline{f}$ such that
\begin{equation}
    \mathcal{T}(f) = \left( \mathcal{T}_{s} \circ \mathcal{T}_{\varphi}\right) \left(f\right) \equiv \overline{f}.
\end{equation}
In the azimuthal direction, it is sufficient to truncate the Fourier expansion to reduce the dimension of the grid while preserving the $[0, 2\pi)$ periodicity. 
The literature contains many examples of ML models that have been trained on doubly-periodic \citep{maulik2019subgrid,guan2022stable} or triply-periodic \citep{sirignano2020dpm,frezat2021physical} datasets. In addition to mode truncation ---also called sharp spectral filter--- one can apply any filter that commutes with partial derivatives. 
Some popular examples include the Gaussian or the Box filter \citep{zhou2019subgrid}. Here we retain only $\overline{N_{m}} \leq N_m$ Fourier modes in the azimuthal direction,
\begin{equation}
    \mathcal{T}_{\varphi}[f(s, \varphi_{k})] = \sum_{m = -\overline{N_m}}^{\overline{N_m}}f_{m}(s) \, e^{\mathrm{ i } m \varphi_{k}}.
\end{equation}
In the radial direction, truncating the Chebyshev series to $\overline{N_{s}} \leq N_s $ polynomials is not sufficient, since the Chebyshev polynomials $T_{n}$ are not boundary-preserving by construction. Instead, we truncate coefficients from a Galerkin basis $\chi_{n}(x_{k})$ that satisfies the appropriate boundary conditions,
\begin{equation}
    \mathcal{T}_{s}[f_{m}(s_{k})] = \sum_{n = 0}^{\overline{N_{s}} - N_{\chi}} f_{mn}^{\mathcal{G}}\, \chi_{n}(x_{k}),
\end{equation}
where $f_{mn}^{\mathcal{G}}$ are the Galerkin coefficients and $N_{\chi}$ is the number of boundary conditions fulfilled by the basis.
In practice, the zonal flow $\langle \overline{u_\varphi} \rangle_{\varphi}$ is expanded onto the following Galerkin basis that satisfies Dirichlet boundary conditions,
\begin{align}
    \chi_{n}(x_{k}) &= T_{n + 2}(x_{k}) - T_{n}(x_{k}), \\
    \chi(\pm 1) &= 0, \\
    N_{\chi} &= 2,
\end{align}
where $x_k$ corresponds to the Gauss--Lobatto grid points (see \ref{eq:GL_grid} in Appendix~\ref{app:numerical-discretisation}).
This decomposition is not the only one that matches Dirichlet boundary conditions on both sides but it appears to be the most commonly employed \citep{shen1995efficient}. For $\overline{\psi}$, we adopt the following Galerkin basis, which satisfies both Dirichlet and Neumann boundary conditions  \citep[e.g.][]{julien2009efficient},
\begin{align}
    \chi_{n}(x_{k}) &= T_{n}(x_{k}) - \frac{2 (n + 2)}{n + 3}T_{n + 2}(x_{k}) + \frac{n + 1}{n + 3}T_{n + 4}(x_{k}), \\
    \chi(\pm 1) &= \chi'(\pm 1) = 0, \\
    N_{\chi} &= 4.
\end{align}
To enforce the incompressibility constraint, we only coarse-grain the streamfunction $\overline{\psi}$ and the zonal velocity $\langle \overline{u_\varphi} \rangle_\varphi$ and then compute the corresponding reduced velocity $\overline{\bm{u}}$ using \eqref{eq:us-from-psi}--\eqref{eq:uphi-from-psi} and vorticity $\overline{\omega}$ using \eqref{eq:omega-from-psi}. 
We illustrate the coarse-graining process using Galerkin and Chebyshev coefficients truncation for an analytical function in Figure~\ref{fig:coarse-graining}. We can observe in the bottom-right 
inset that, unlike the Chebyshev basis, the Galerkin basis preserves the boundary conditions $z(\pm 1) = 0$ during the coarse-graining process.

\subsection{Implicit subgrid terms using online learning}
From a ML point-of-view, the subgrid-scale problem is most of the time formulated as a supervised learning problem,
\begin{equation}
    \operatorname*{arg\,min}_\theta \, L \left[ \tau, \mathcal{M}(\overline{\mathbf{z}}; \theta) \right] \label{eq:problem-offline}
\end{equation}
where $\theta$ are the trainable parameters of model $\mathcal{M}$, minimizing a given loss function $L$. With this so-called ``offline'' approach \citep{sanderse2025scientific}, we need to build a dataset $\mathbb{D}$ of tuples (input, target) such that $\mathbb{D} \coloneq \{\overline{\mathbf{z}}\} \rightarrow \{\tau\}$. The SGS term $\tau$ comes from the coarse-graining of the original equations \eqref{eq:sph-qg-omega} and \eqref{eq:sph-qg-uphi-axisymmetric}. Within a learning framework, we aim to approximate the following SGS terms,
\begin{align}
  \tau = 
  \begin{bmatrix}
  \tau_{\omega} \\ 
  \tau_{\langle u_{\varphi} \rangle_{\varphi}}
  \end{bmatrix} = 
  \begin{bmatrix}
  \bm{\nabla} \bcdot \left[ \mathcal{T}(\bm{u})\mathcal{T}(\omega) - \mathcal{T}(\bm{u} \omega) \right] \\
  \left\langle \mathcal{T}(u_{s})\mathcal{T}(\omega) - \mathcal{T}(u_{s} \omega) \right\rangle_{\varphi}
  \end{bmatrix} \approx 
  \mathcal{M}(\overline{\mathbf{z}}; \theta).
  \label{eq:sgs-term}
\end{align}
This definition, however, is valid only under the assumption that $\mathcal{T}$ commutes with partial derivatives \citep[chapter~2.2]{sagaut2006large}. For the annular geometry, the Gauss--Lobatto grid in the radial direction is not homogeneous and is thus responsible for the non-commutativity of the filtering and differentiation operators. This introduces a commutation term in the governing equations \citep{ghosal1995basic}. In practice, this term is however often ignored leading to a commutation error that is not well understood, in particular during the coarse-graining process \citep[see the investigation][]{yalla2021effects}.
Recently, \citet{jakhar2024learning} trained a model for Rayleigh--Bénard convection based on a similar Fourier--Chebyshev spatial discretization; they applied filtering in the periodic direction only, thereby avoiding the commutation error. 
With regard to the QG turbulence problem of interest here, let us briefly state that experiments based on the ``offline'' approach were systematically unsuccessful, in the sense that the trained SGS models always led to unstable simulations. We will consequently not discuss the ``offline'' approach further. 
The main objective of model $\mathcal{M}$ is to lead to an accurate evolution of the coarse variables $\overline{\mathbf{z}}(t)$ with respect to that of the direct variables $\mathbf{z}(t)$. A more realistic supervised learning problem would write
\begin{equation}
    \operatorname*{arg\,min}_\theta \, L \left[ \mathbf{z}(t), \overline{\mathbf{z}}(t) \right],
    \label{eq:problem-online}
\end{equation}
where $\overline{\mathbf{z}}(t)$ is bound to the subgrid model $\mathcal{M}$ by the flow operator $\gamma_{\theta}$,
\begin{equation}
    \overline{\mathbf{z}}(t) = \gamma_{\theta}(\overline{\mathbf{z}}, t_{0}, t) = \overline{\mathbf{z}}(t_{0}) + \int_{t_{0}}^{t} \left[ \mathcal{L}\overline{\mathbf{z}}\left(t'\right) + \mathcal{N}\left(\overline{\mathbf{z}}, t'\right) + \mathcal{M}\left(\overline{\mathbf{z}}(t'); \theta\right) \right] \, \mathrm{d}t' \label{eq:flow-operator}
\end{equation}
that advances the system \eqref{eq:sph-qg-omega-coarse} and \eqref{eq:sph-qg-uphi-axisymmetric-coarse} from time $t_{0}$ to time $t$, while optimizing for the parameters $\theta$ such that the loss function $L$ in \eqref{eq:problem-online} is minimized. 
In its discrete form, the evolution operator $\gamma_{\theta}(\overline{\mathbf{z}}, t_{0}, t)$ corresponds to solving multiple sub-stages of the IMEX integrator. 
As the model $\mathcal{M}$ contains nonlinearities, it is treated explicitly as an additive source term, which results in practice in three calls per time step within the BPR353 IMEX scheme. 
It is clear that the flow operator \eqref{eq:flow-operator} in the supervised learning problem \eqref{eq:problem-online} does not explicitly require knowledge of the subgrid term $\tau$.
This ``online'' formulation \eqref{eq:problem-online} is often pointed out as more stable \citep{frezat2022posteriori,sanderse2025scientific} compared to the ``offline'' formulation \eqref{eq:problem-offline}. 
In addition, we propose here that the model $\mathcal{M}$ can implicitly learn to correct other error sources such as, e.g., the commutation error due to the radial coarse-graining. This formulation also allows some flexibility in prescription of the loss function $L$. Here, we define a kinetic energy mismatch metric,
\begin{equation}
    L \left[ \mathbf{z}(t), \overline{\mathbf{z}}(t) \right] = \pi \sideset{}{'}\sum_{m = 0}^{\overline{N_{m}}} \int_{s_{i}}^{s_{o}} \left( |\overline{u_{s_{m}}}(t) - u_{s_{m}}(t)|^{2} + |\overline{u_{\varphi_{m}}}(t) - u_{\varphi_{m}}(t)|^{2} \right) s \, \mathrm{d} s.
\end{equation}
Numerically, solving \eqref{eq:problem-online} is performed using a flavor of stochastic gradient descent and thus requires the gradient of the loss function $\bm{\nabla}_{\theta} L$. Expanding using the flow operator \eqref{eq:flow-operator} gives
\begin{equation}
    \frac{\partial L}{\partial \theta} = \left( \frac{\partial L}{\partial \gamma_{\theta}} \right)^{\mathsf{T}} \left[ \int_{t_{0}}^{t} 
    \left(
    \frac{\partial \mathcal{L}\overline{\mathbf{z}}(t')}{\partial \theta} + \frac{\partial \mathcal{N}(\overline{\mathbf{z}}, t')}{\partial \theta} + \frac{\partial \mathcal{M}(\overline{\mathbf{z}}(t'); \theta)}{\partial \theta} \right) \, \mathrm{d}t' \right].
    \label{eq:gradient-loss}
\end{equation}
It appears that optimizing for the parameters $\theta$ using the ``online'' formulation requires the adjoint of the model $\mathcal{M}$ and the implicit $\mathcal{L}$ and explicit $\mathcal{N}$ forward solver terms.
There exists many strategies to obtain these quantities, either exact or approximate \citep{ouala2024online}, which are not without drawbacks.
Obtaining the exact adjoint can be done either by an explicit derivation, or by rewriting the solver using a differentiable framework, which can represent a substantial undertaking. On the other hand, approximate approaches are often computationally prohibitive, and can introduce sources of error in the optimization problem \citep{frezat2026gradient}.
In our case, the numerical method is simple enough for us to implement a solver from scratch using the JAX library \citep{jax2018github}. The library algorithmically provides us with exact derivatives using automatic differentiation \citep{baydin2018automatic}.

\subsection{Learning setup \label{sec:learning_setup}}
In a classical ``offline'' setup, data are generated such that samples are not correlated. For the ``online'' setup, we need continuous sub-trajectories from the DNS, spaced at time intervals matching those of the reduced resolution. In practice, it is easier to work with constant time steps. 
Let us denote stable direct-resolution and reduced-resolution time steps $\delta t$ and $\overline{\delta t}$, respectively. The reduced simulation is defined on a coarser grid, and thus we have $\overline{\delta t} > \delta t$. Moreover, for simplicity, we assume that $\overline{\delta t}$ is related to $\delta t$ by an integer factor $\alpha_{\text{coarse}}$ such that $\overline{\delta t} = \alpha_{\text{coarse}} \delta t$. Finally, we want to generate a dataset mapping initial sub-trajectory conditions $\overline{\mathbf{z}}(t_{0})$ to high-resolution (or coarse-grained high-resolution) variables that align with the reduced-resolution time step $\overline{\delta t}$,
\begin{equation}
    \mathbb{D} \coloneq \{ \overline{\mathbf{z}}(t_{0}) \} \rightarrow \{ \mathbf{z}(t_{0} + s\overline{\delta t}) \}_{s \in [1, N_{\mathrm{steps}}]},
\end{equation}
where $N_{\mathrm{steps}}$ is the number of discrete time steps required to integrate from $t_{0}$ to $t$ in \eqref{eq:flow-operator}. The value of $N_{\mathrm{steps}}$ is often limited by the memory available on the device, since automatic differentiation keeps track of the gradient graph built from any operation applied in the loss function $L$. Although it is possible to increase this limit with checkpointing \citep{sapienza2024differentiable}, the width of the time interval over which adjoint optimisation is performed is constrained by the intrinsic limit of predictability of the system due to turbulence (see the discussion on training timescales in Sec. \ref{sec:training-timescales} below). For consistency, the size of the dataset is determined by the dynamics of a given configuration, defined in this study by the geometry of the container, the forcing ${\mathcal F}$, and the value of the Ekman number $E$. In practice, we build datasets that span one eddy turnover time for each of the three configurations that will be detailed in the next section. Models are trained using the AdamW algorithm \citep{loshchilov2018decoupled} with a fixed learning rate of $10^{-4}$.

\subsection{Spectral-space architecture}
Model $\mathcal{M}$ should operate in spectral space, since it is meant to integrate smoothly in an existing pseudo-spectral solver. Operating in spectral space avoids potential back and forth transforms between spectral space and grid space, but demands complex-valued neural networks \citep{hirose2006complex,trabelsi2018deep}. 
In addition, model $\mathcal{M}$ should have an inference time that does not exceed that of the calculation of the linear and non-linear terms of \eqref{eq:sph-qg-omega-coarse} and \eqref{eq:sph-qg-uphi-axisymmetric-coarse}. We take inspiration from modern convolutional blocks, also called ConvNext, which have been shown to compete favorably with transformers in terms of accuracy and scalability \citep{liu2022convnet}. We propose a low-complexity variant of the described ConvNext architectures in order to reduce the computational cost at inference. 
\begin{figure}
    \centering
    \includegraphics[width=0.55\linewidth]{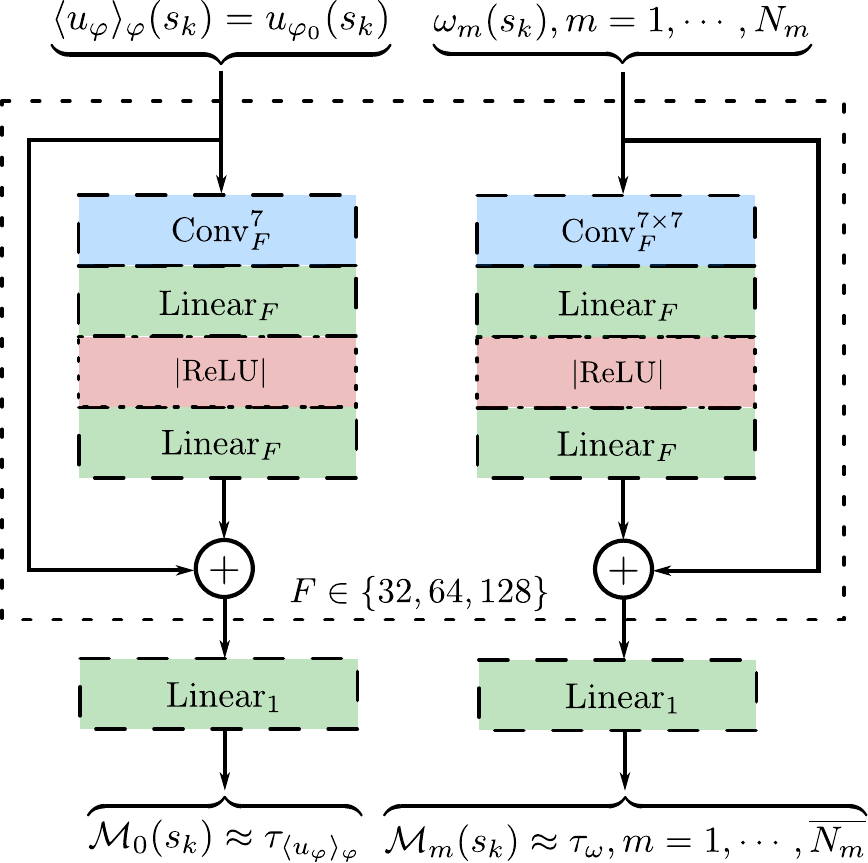}
    \caption{Illustration of the spectral-space architecture. The architecture is split into two paths for the axisymmetric velocity and the vorticity equations. The model applies three (residual) blocks of increasing number of features $F \in \{32, 64, 128\}$ containing each a conv layer, a linear layer and a non-linear mod ReLU, or $|\mathrm{ReLU}|(z) = \mathrm{ReLU}(z + 1) z / |z|$ activation \citep{arjovsky2016unitary} followed by a Linear layer. The last Linear layer projects the high-dimensional features $F$ into a single feature. The total number of trainable parameters for this architecture is approximately 666 000. \label{fig:spectral-architecture}}
\end{figure}
The complexity of this model, i.e., size of kernels, number of layers and features has been selected by trial and error.
The overall model $\mathcal{M}$ is also divided into two paths that produce subgrid corrections to \eqref{eq:sph-qg-omega-coarse} and \eqref{eq:sph-qg-uphi-axisymmetric-coarse}, respectively. An illustration with the inner details of the spectral-space architecture is shown in Figure~\ref{fig:spectral-architecture}.
The left path is applied to $\langle u_{\varphi} \rangle_{\varphi}$, namely the axisymmetric $m = 0$ mode of the azimuthal velocity, while the right path is applied to $\omega$ for all non-axisymmetric modes $m > 0$. In practice, the left path takes a one-dimensional (1D) input of size $\overline{N_{s}}$, while the right path takes a 2D input of size $(\overline{N_{m}} - 1, \overline{N_{s}})$. 
Internally, each trainable layer is real-valued operating on complex-valued input, with backpropagation supported by Wirtinger derivation rules \citep[e.g.][]{hjorungnes2007complex}, i.e., real and imaginary parts are treated separately.

\section{Results \label{sec:results}}
To evaluate the capabilities of the learned subgrid models, we explore three different configurations that yield a turbulent quasi-geostrophic regime. Numerically, the maximum grid resolution that can fit in GPU memory is limited by the imprint of the dense collocation matrices \eqref{appeq:omega-psi-system}; consequently, we use $N_s = 321$ for each configuration in the radial direction with $N_m = 400$ for configuration (i) and $N_m = 480$ for configurations (ii-iii) in the azimuthal direction.
For the coarse-grained resolution, the original grids have been truncated to $(\overline{N_s}, \overline{N_m}) = (\lceil N_s/\alpha_\text{coarse} \rceil, N_m/\alpha_\text{coarse})$ with $\alpha_\text{coarse} = 5$, which corresponds to a fivefold ratio compared to the DNS grid. At this resolution, the Ekman boundary layers are barely resolved since they only contain approximately five grid points \citep[see][]{shishkina2010boundary}. The subgrid-scale problem becomes significantly more difficult with a 10-fold ratio grid; in that case, the Ekman layers are definitively not resolved, and we did not explore the 10--times coarser resolutions further. 
For each physical configuration, the transient regime is only computed using DNS. Dataset generation and evaluation start when a statistically steady state has been reached.
\begin{figure}
  \centerline{
  \includegraphics[width=.45\linewidth]{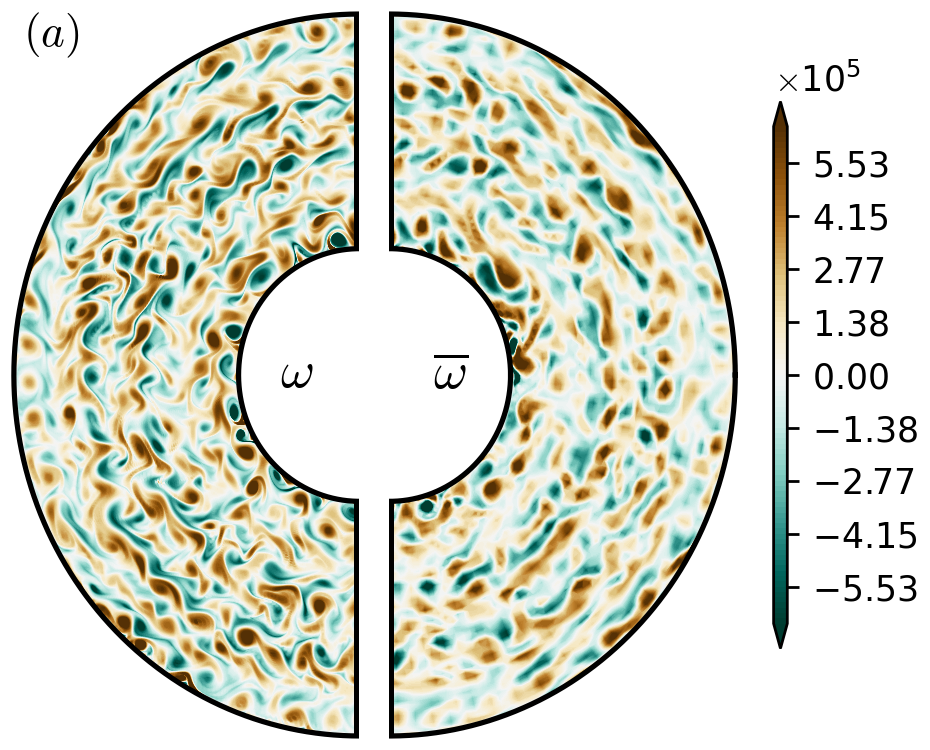}
  \includegraphics[width=.45\linewidth]{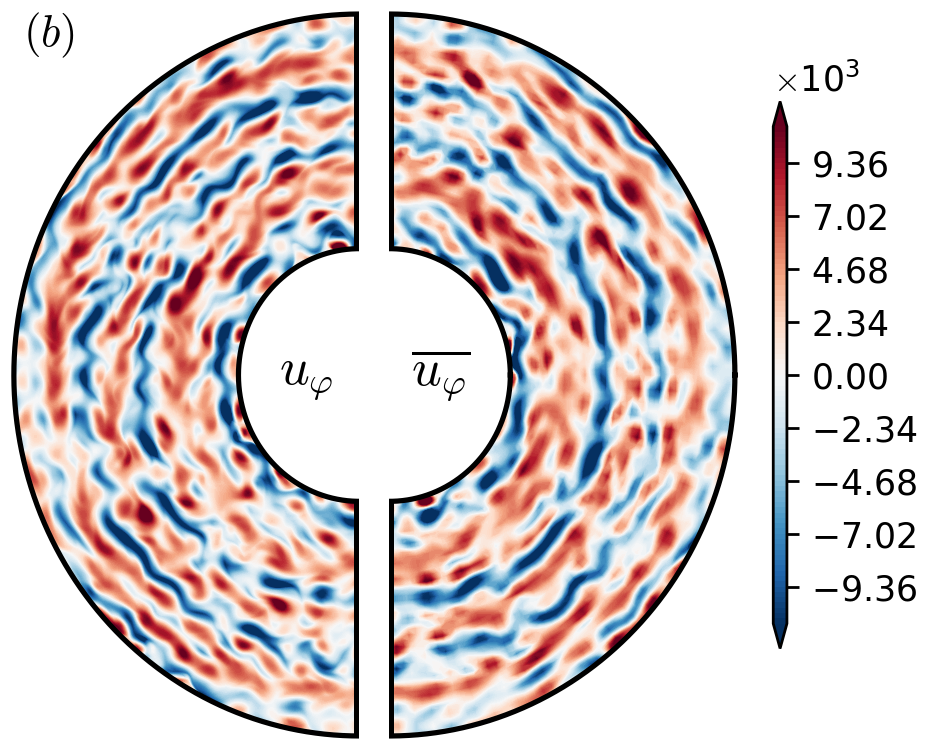}}
  \centerline{
  \includegraphics[width=.45\linewidth]{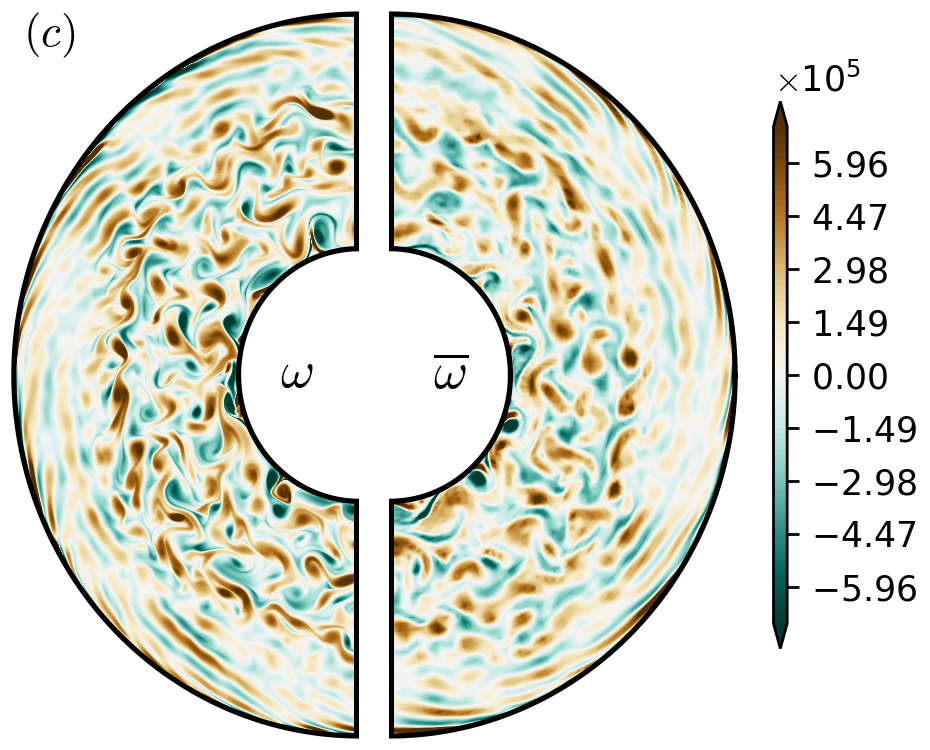}
  \includegraphics[width=.45\linewidth]{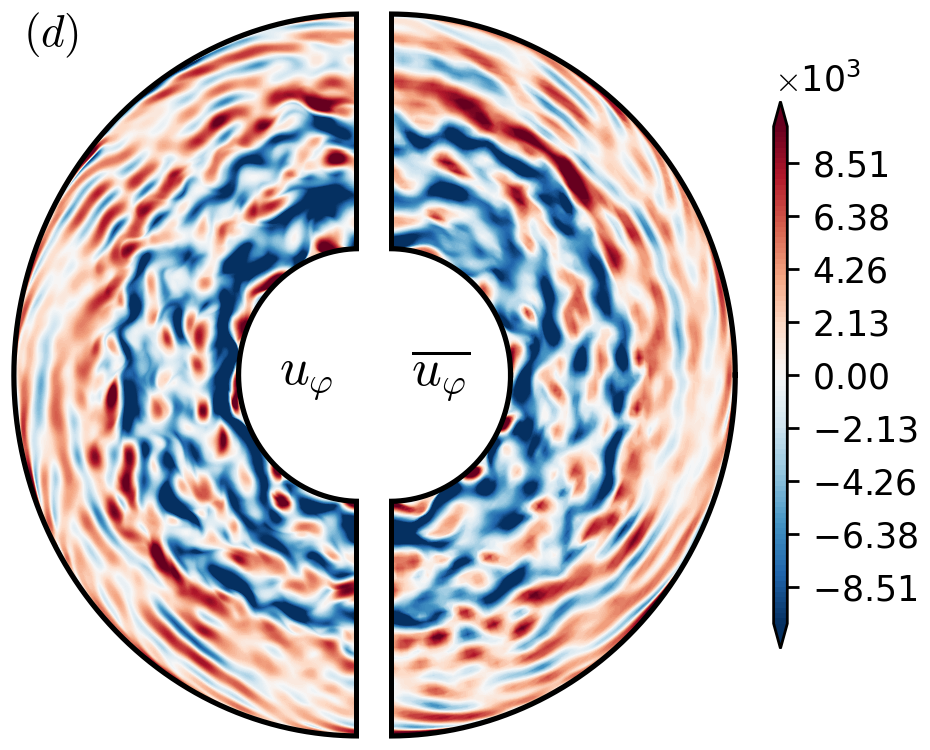}}
  \centerline{
  \includegraphics[width=.45\linewidth]{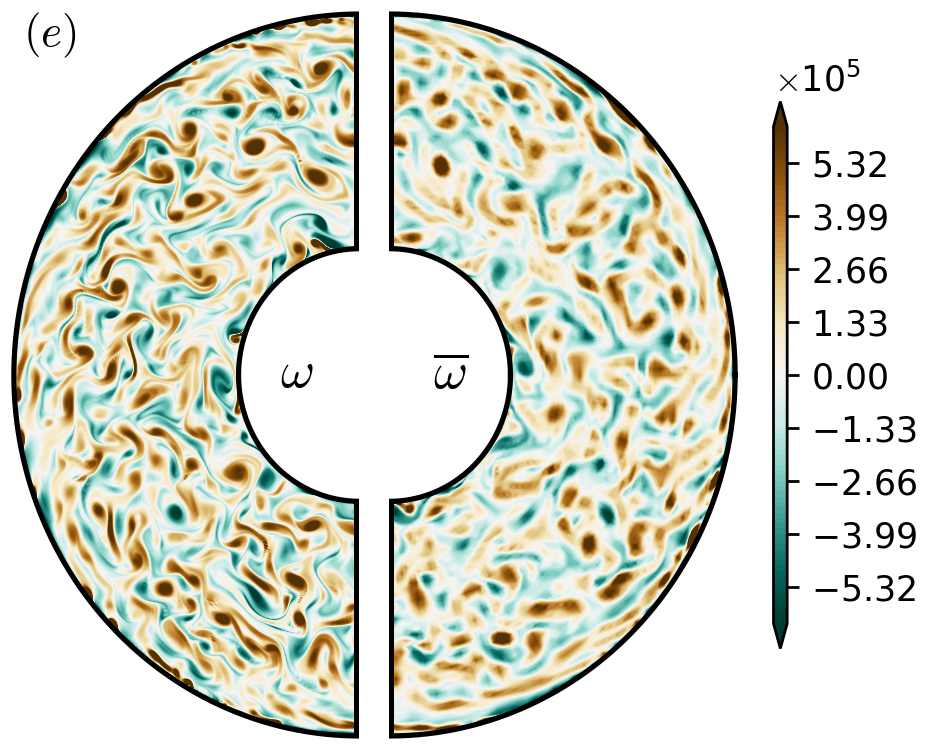}
  \includegraphics[width=.45\linewidth]{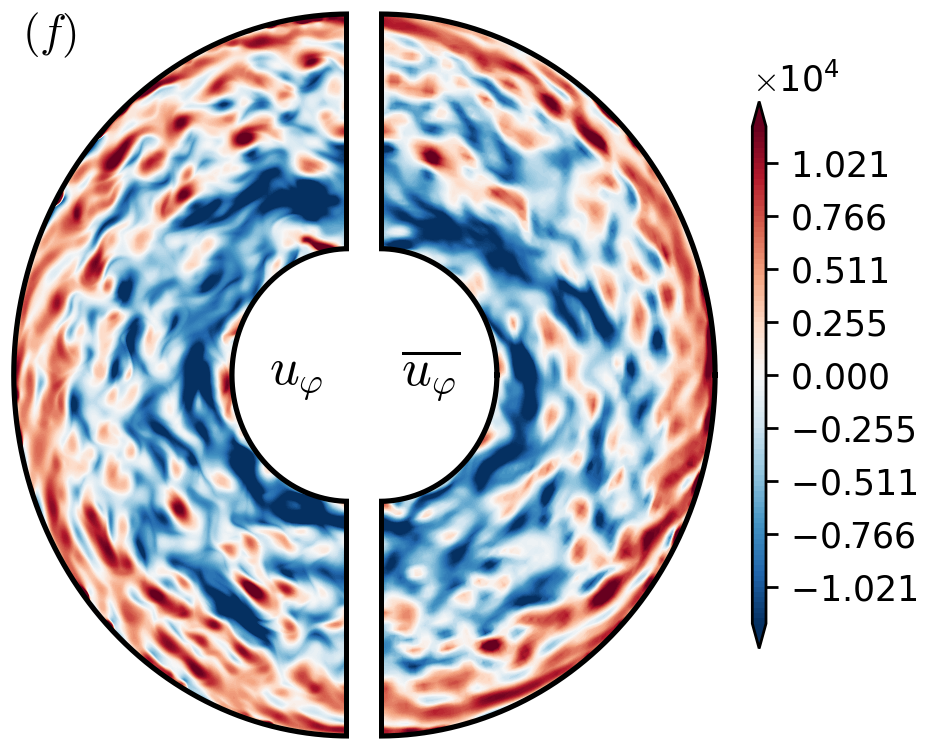}}
  \caption{Vorticity $\omega$ (left) and azimuthal velocity $u_{\varphi}$ (right) from DNS for the three configurations. (a-b): configuration (i), exponential container with $E = 2 \times 10^{-7}$ and $\beta = -1 / s_{o}$. (c-d): configuration (ii), spherical shell with $E = 3 \times 10^{-7}$. (e-f): configuration (iii), spherical shell with $E = 10^{-6}$. In each panel, the left half shows the full-resolution fields, while the right half corresponds to their coarse-grained counterparts after Fourier--Galerkin truncation. See text for details. \label{fig:configurations-fields}}
\end{figure}
The three configurations share a common forcing amplitude $a_{\mathcal{F}}= 2 \times 10^{10}$. To investigate the influence of rotation and shape of the 3D container on the turbulent flow, we vary both the Ekman number and the $\beta(s)$ profile. 
\begin{table}
    \begin{center}
    \begin{tabular}{r l c c c c c c c}
    & Geometry & $(N_s, N_m)$ & $E$ & $a_{\mathcal{F}}$ & $\delta t$ & $\turnover$ & $\lambda$ & $\mathrm{dim}(\mathbb{D})$ \\
    \hline
    (i) & Exp. ($\beta = -1 / s_{o}$) & $(321, 400)$ & $2 \times 10^{-7}$ & $2 \times 10^{10}$ & $5 \times 10^{-8}$ & $1.64 \times 10^{-4}$ & $2.1 \times 10^{4}$ & 675 \\ 
    (ii) & Spherical & $(321, 480)$ & $3 \times 10^{-7}$ & $2 \times 10^{10}$ & $4 \times 10^{-8}$ & $1.55 \times 10^{-4}$ & $2.7 \times 10^{4}$ & 800 \\
    (iii) & Spherical & $(321, 480)$ & $1 \times 10^{-6}$ & $2 \times 10^{10}$ & $4 \times 10^{-8}$ & $1.32 \times 10^{-4}$ & $3.1 \times 10^{4}$ & 675 
    \end{tabular}
    \caption{Numerical model parameters: grid resolution in the azimuthal and radial directions $(N_s, N_m)$, Ekman number $E$, forcing amplitude $a_{\mathcal{F}}$ and steady-state fixed time step size $\delta t$. Note that a smaller time step size might be required to ensure stability during the transient. The turnover time $\turnover$ and the growth rate $\lambda$ provide statistical information about the dynamics of the simulation. These time-related statistics are used to estimate the size of the dataset $\mathrm{dim}(\mathbb{D})$.}
    \label{tab:parameters-configurations}
    \end{center}
\end{table}
In the remainder of this study, we will refer to the three setups summarized in Table~\ref{tab:parameters-configurations} and illustrated in Figure~\ref{fig:configurations-fields} using the following convention: 
\begin{enumerate}[i]
\item \textbf{Exponential container with $\beta = -1/ s_{o}$ and $E = 2\times 10^{-7}$} (Figure~\ref{fig:configurations-fields}, top row). Due to the dominant role of rotation, we observe the formation of zonal jets of alternating directions, which yield concentric rings of vorticity. As $\beta$ is constant, a similar dynamics is observed throughout the domain. Fluctuations about the mean flow have a  magnitude comparable to that of zonal jets. This configuration leads to multistable solutions, as the number and position of jets vary across different realizations \citep[as in e.g. Fig.~6 of][]{lemasquerier2023zonal}.
\item\textbf{Spherical container with $E = 3\times 10^{-7}$} (Figure~\ref{fig:configurations-fields}, middle row). This configuration exhibits two dynamical regions. In the inner part of the annulus, $|\beta|$ is minimum (Table~\ref{tab:geometric-quantities}), and the dynamics is dominated by turbulence with a predominance of vortices, while the external part of the annulus shows spiralling structures typical of Rossby waves. This segregated dynamics is typical of the flow observed in spherical containers, in which $|\beta|$ increases quickly outwards  \citep[see e.g. Fig.~5 in][]{barrois2022comparison}.  
Segregation follows from the local competition between advection-dominated processes, whose timescale is the local turnover time, and propagation-dominated processes, whose timescale is the Rossby wave period. The transition between the two regions occurs when those two timescales are comparable \citep{guervilly2017multiple}. 
\item\textbf{Spherical container with $E = 10^{-6}$} (Figure~\ref{fig:configurations-fields}, bottom row). Configuration (iii) is configuration (ii) with an increased Ekman number. Advective processes therefore prevail in most of the fluid volume, confining Rossby waves to a thin fluid region close to the outer boundary. The pair of prograde and retrograde jets are stronger in magnitude compared with that of configuration~(ii).
\end{enumerate}

We adopt the following averaging definitions to compute diagnostics and evaluation metrics,
\begin{equation}
    \langle f \rangle_{\varphi} = \frac{1}{2 \pi} \int_{0}^{2 \pi} f \, \mathrm{d}\varphi, \quad \langle f \rangle_{s} = \frac{2}{s_o^2 - s_i^2} \int_{s_{i}}^{s_{o}} f \, s\, \mathrm{d}s, \quad \langle f \rangle = \langle f \rangle_{\varphi,s},
    \label{eq:def_average}
\end{equation}
where $\langle \cdot \rangle_{\varphi}$ corresponds to azimuthal averaging, $\langle \cdot \rangle_{s}$ to radial averaging and $\langle \cdot \rangle$ to domain (surface) average, i.e., in both azimuthal and radial directions.

The total surface kinetic energy density $E_{T}$ can be decomposed into a zonal component $E_{Z}$ for the axisymmetric flow and a residual component $E_{R}$
\begin{equation}
    E_{T} = \frac{1}{2} \langle \bm{u}^{2} \rangle = E_Z + E_R = \frac{1}{2} \langle u_{\varphi_{0}}^{2} \rangle_{s} + \sum_{m = 1}^{N_{m}} \langle u_{s_{m}}^{2} \rangle_{s} + \langle u_{\varphi_{m}}^{2} \rangle_{s}\,.
    \label{eq:ET}
\end{equation}
The Reynolds and Rossby numbers are accordingly defined by
\begin{equation}
    Re = \sqrt{2 E_T}, \quad Ro = Re E.
\end{equation}
We are also interested in quantities that are more sensitive to the small scales of the flow, closer to where subgrid-scale transfers occur, such as the total enstrophy density
\begin{equation}
    \enstrophy = \frac{1}{2} \langle \left(\bm{\nabla} \times \bm{u}\right)^{2} \rangle = \frac{1}{2} \langle \omega^{2} \rangle.
\end{equation}
From the ratio of kinetic energy and enstrophy, one introduces the viscous dissipation lengthscale
\begin{equation}
    \dissls = \sqrt{\frac{E_T}{\enstrophy}}.
    \label{eq:ldiss}
\end{equation}
The numerical simulations are also evaluated in terms of their power balance, which is obtained by the inner product of the Navier--Stokes equations with $\bm{u}$, followed by spatial averaging. In our setups, turbulence is driven by the power input 
\begin{equation}
    \power = \langle \bm{f} \bcdot \bm{u}\rangle,
\end{equation}
where $\bm{f}$ stands for the driving force that sustains the forcing pattern shown in Figure~\ref{fig:qg-forcing}. An analytical expression of $\power$ is derived in Appendix~\ref{sec:power-input}. Kinetic energy is dissipated both by Ekman friction and bulk viscous dissipation. 
Even if there is no analytical expression for the power loss by Ekman friction, denoted by $\mathcal{D}_\Upsilon$, we can use the time average energy budget to retrieve its time average value. The instantaneous power budget indeed reads 
\begin{equation}
    \frac{\mathrm{d} \langle E_T \rangle}{\mathrm{d}t} = \power - \langle \omega^2\rangle - \mathcal{D}_\Upsilon, 
    \label{eq:power_balance}
\end{equation}
where the second-term on the right-hand side corresponds to the bulk viscous dissipation. 
Upon time averaging, denoted by $\langle \cdot \rangle_t$, we get 
\begin{equation}
    0 = \langle \power \rangle_t -  \langle \omega^2 \rangle_t - \langle \mathcal{D}_\Upsilon \rangle_t, 
\end{equation}
which allows us to characterize the relative fraction of energy dissipated through Ekman friction
\begin{equation}
    \rfric \equiv \frac{\langle \mathcal{D}_\Upsilon \rangle_t}{\langle \mathcal{D}_\Upsilon \rangle_t  + \langle \omega^2\rangle_t}.
    \label{eq:tavg_eqbalance}
\end{equation}

\subsection{Training timescales \label{sec:training-timescales}}
Our goal in this study is not to evaluate the ability of a given model to generalize over different initial conditions, parameters, or grid resolutions. Instead, we explore the possibility of training a model to extend the initial integration of a DNS  at reduced numerical cost over a long temporal horizon. As exposed in Sec. \ref{sec:learning_setup}, we build datasets from the DNS that span one turnover time $\turnover = 1/Re$ for each configuration and proceed with reduced simulations for $100$~turnover times.
The number of DNS samples used for online learning is thus $\mathrm{dim}(\mathbb{D}) \approx \turnover / \overline{\delta t}$, where the coarse time step size $\overline{\delta t}$ is adjusted considering grid ratios between the direct and coarse-grained resolutions. Here, since the time step is limited by the azimuthal part of the advection term, we adopt $\overline{\delta t} = 5 \delta t$, due to the five--fold coarsening of the azimuthal grid. Table~\ref{tab:parameters-configurations} indicates that the three configurations have similar values for $\turnover$, configuration (iii) being the most turbulent; it has the largest Ekman number, while all configurations share the same forcing amplitude $a_{\mathcal{F}}$. The size of the datasets ranges from $675$ samples for configurations~(i) and (iii) to $850$ samples for configuration~(ii). The number of consecutive samples $N_{\mathrm{steps}}$ used for the online learning strategy is set by a practical constraint: accumulating gradients in the loss computation \eqref{eq:gradient-loss} takes up GPU memory, which, in practice, limits us to $N_{\mathrm{steps}} = 25$ time steps separated by a constant $\overline{\delta t}$. In any event, since we study deterministically chaotic dynamics, we must check that this hardware-imposed training timescale  is small compared with the decorrelation time of each configuration, which governed by the nonlinear flow operator in \eqref{eq:flow-operator}.
In order to quantify the divergence of the three configurations over a sub-trajectory timespan, we consider for each configuration a reference vorticity $\omega_{\mathrm{ref}}$ which has reached statistical equilibrium, and which we perturb by a tiny amount at some arbitrary time $t_p$.
\begin{figure}
  \includegraphics[width=.95\linewidth]{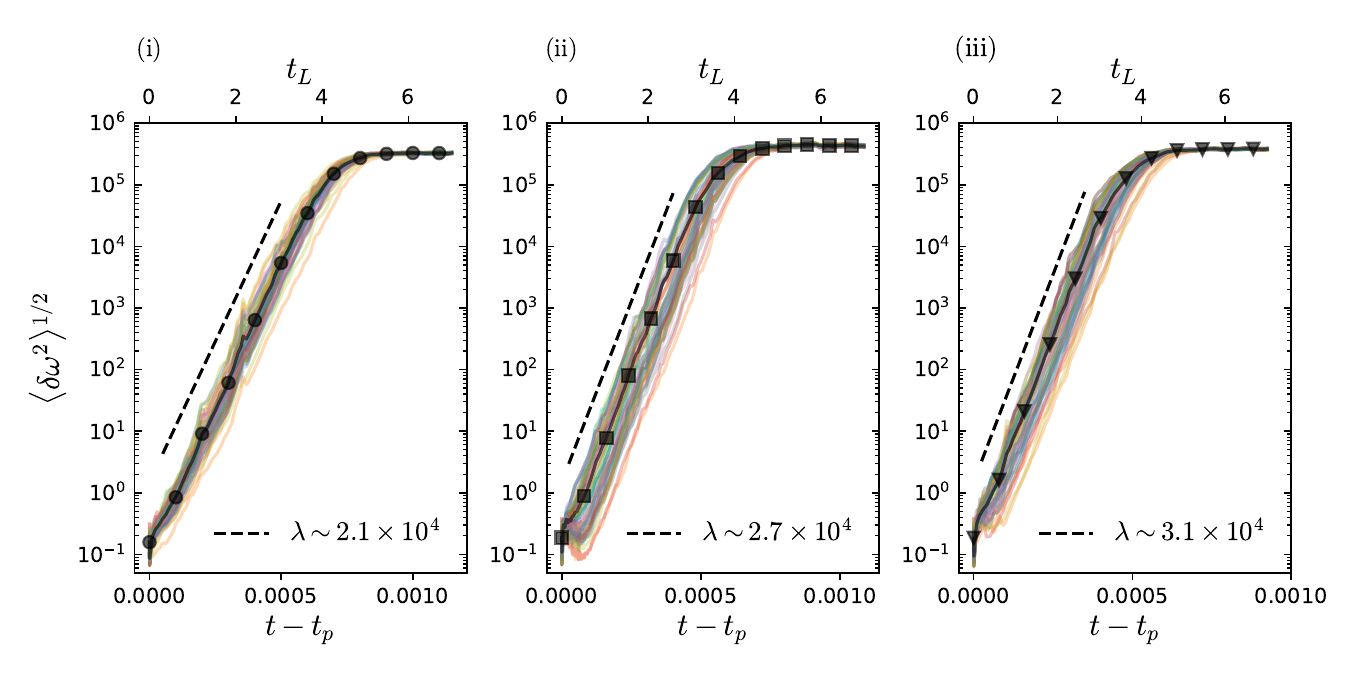}
  \caption{Evolution of the deviation (based on the root mean squared vorticity error) for ensembles of perturbed simulations for the configurations (i), (ii) and (iii).
  The perturbation is set at $t = t_p$ as a single randomly localized Gaussian source similar to those appearing in \eqref{eq:forcing}, activated with an amplitude  $10^{-10} a_\mathcal{F}$. In each panel, the dashed line shows exponential growth for a growth rate $\lambda$ obtained by least-squares fit. \label{fig:decorrelations}}
\end{figure}
If $\omega_p$ refers to the subsequent evolution of the perturbed solution, the growth of the difference $\delta \omega = \omega_p - \omega_{\mathrm{ref}}$ is initially exponential, with a growth rate $\lambda$ such that
\begin{equation}
 \langle \delta \omega^2 \rangle ^{1/2} \propto \exp \lambda (t-t_p). 
\end{equation}
For each configuration, we define an ensemble of perturbations at $t_p$  by activating, for each ensemble member, a random pump in the collection given in \eqref{eq:forcing}, at a modest amplitude of $10^{-10} a_\mathcal{F} = 2$. We show in Figure~\ref{fig:decorrelations} the evolution of the norm of the difference between the reference solution and the ensemble of perturbed trajectories. In each panel, a phase of exponential growth if followed by a saturated phase. 
Dashed lines show exponential growths for values of $\lambda$ obtained by least-squares, which are listed in Table~\ref{tab:parameters-configurations} and correspond to timescales  $\lambda^{-1}$ in the range $0.2-0.3$~$\turnover$. Figure~\ref{fig:decorrelations} shows that each configuration is completely decorrelated at $t - t_p \approx 5 \turnover \approx 7 \times 10^{-4}$, which is more than 100 times larger than the sub-trajectory timespan for learning of $25\, \overline{\delta t} \geq 5\times 10^{-6}$. In other words, the timespan chosen for online learning represents $\mathcal{O}(1)$\% of the physical decorrelation time for all three configurations, which ensures that the gradient-based optimization is meaningful.

\subsection{Alternative models}
We henceforth denote by $\tau \equiv \mathcal{M}$ the model trained with the implicit ``online'' scheme described above and $\tau \equiv 0$ the absence of any correction, i.e., a simulation at reduced resolution (also referred to as under-resolved simulation hereafter). 
We assess the performance of $\tau \equiv \mathcal{M}$ against a standard hyperdiffusion, denoted as $\tau \equiv d(m)$. 
In practice here, hyperdiffusion only acts along the azimuthal direction such that the viscous terms in \eqref{eq:sph-qg-omega-coarse}--\eqref{eq:sph-qg-uphi-axisymmetric-coarse} are multiplied in spectral space by a factor 
\begin{equation}
    d(m) = a_d^{m - m_{d}},\, m_{d} \leq m < \overline{N_{m}},
\end{equation}
where $a_d$ is the hyperdiffusivity amplitude and $m_{d}$ is the wavenumber beyond which hyperdiffusivity operates \citep{nataf2015turbulence}. Following the prescriptions by \citet{davy2026effect}, $m_{d}$ is set to two times the wavenumber of the peak of the radial velocity spectrum $|u_{s_m}|^2$. This corresponds to $m_{d} \approx 56$ for configuration (i) and $m_{d} \approx 24$ for configurations (ii) and (iii).
The amplitude $a_d$ is calibrated by trial and error such that $a_d$ suppresses energy pile-up at small-scales, with values ranging from $a_d = 1.10$ for configuration (i) to $a_d = 1.02$ for configuration (ii,iii).

We also examine the performance of the Leith closure model \citep{leith1996stochastic}, denoted as $\tau \equiv \nu_\text{L}$, a standard eddy-viscosity approach for 2D turbulence. The Leith model acts on the direct enstrophy cascade, increasing the viscosity until the resolved dissipation scale. It is widely adopted in the context of quasi-2D turbulence \citep[e.g.][]{fox2008can} and QG turbulence to model oceanic circulation \citep[e.g][]{bachman2017scale,grooms2023backscatter}. In practice, the Leith model is represented by a space and time dependent eddy-viscosity of the form
\begin{equation}
    \nu_\text{L}(s,\varphi,t) = \text{min} \left\{ \nu_\text{L}^\text{max}(s), \left[ \frac{\Lambda \Delta(s)}{\pi} \right]^3 \sqrt{\left(\dfrac{\partial \omega}{\partial s}\right)^2 +
    \left(\dfrac{1}{s}\dfrac{\partial \omega}{\partial \varphi}\right)^2} \right\}, \ \nu_\text{L}^\text{max}(s)=\dfrac{1}{4}\dfrac{\Delta^2(s)}{\delta t},
\end{equation}
where $\Delta(s) = \min(\Delta_s(s), s\Delta_\varphi)$ is the minimum local grid spacing of the annulus and $\Lambda$ is a dimensionless constant of order unity. Since the eddy-viscosity is handled explicitly by the time stepping scheme, its value is bounded by a ceiling function $\nu_\text{L}^\text{max}(s)$ to ensure numerical stability \citep{grooms2023backscatter}.
The subgrid-scale model then amounts to include the following additional terms into the right hand sides of \ref{eq:sph-qg-omega-coarse} and \ref{eq:sph-qg-uphi-axisymmetric-coarse},
\begin{equation}
    \tau(s, \varphi, t) = 
    \begin{bmatrix} 
        \tau_\omega \\ 
        \tau_{\langle u_\varphi \rangle_\varphi}
    \end{bmatrix} = 
    \begin{bmatrix}
        \bm{\nabla} \bcdot \left( \nu_\text{L} \bm{\nabla} \omega \right) \\
        \bm{\nabla} \bcdot \left( \langle \nu_\text{L} \rangle_\varphi \bm{\nabla} \langle u_\varphi \rangle_\varphi \right)
    \end{bmatrix}.
\end{equation}
The values of $\Lambda$ that provide the best fit to the DNS are $2.0$ for configurations (i) and (iii) and $1.5$ for configuration (ii); these values being close to those adopted by e.g. \citet{wilder2025examining}.
\begin{figure}
  \includegraphics[width=.95\linewidth]{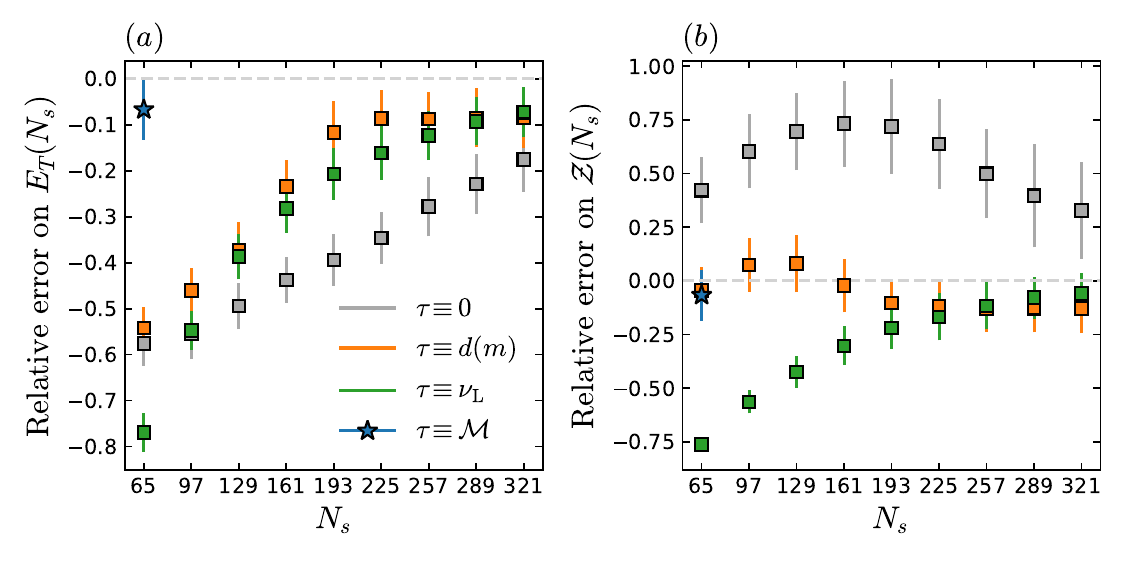}
  \caption{(\textit{a}) Relative error on the kinetic energy density $E_T$ and (\textit{b}) enstrophy density $\enstrophy$ between DNS and different models as a function of their effective radial resolution $N_s$ for configuration (ii). In both panels, positive (negative) values correspond to integrated quantities that exceed (underestimate) the DNS target values. The stars correspond to the resolution of the proposed learned model $\tau \equiv \mathcal{M}$ at the coarsest radial resolution $N_s = 65$. \label{fig:metrics_ns}}
\end{figure}

The aim of the model $\tau \equiv \mathcal{M}$ proposed here is to run at resolutions coarsened in both the periodic azimuthal and bounded radial directions. In contrast, studies that incorporate hyperdiffusion along the periodic horizontal directions usually maintain a resolution comparable to that of the DNS along the bounded vertical axis to preserve the boundary layer structure, while coarsening the resolution in the directions where hyperdiffusion applies \citep[e.g.][]{chen2005large,davy2026effect}. 
To explore the impact of radial coarsening, Fig.~\ref{fig:metrics_ns} evaluates the relative error of the alternative models $\tau \equiv d(m)$ and $\tau \equiv \nu_\text{L}$ as a function of the radial resolution $N_s$ for configuration (ii) in terms of two integrated quantities, $E_T$ and $\enstrophy$. Note that results for configurations (i) and (iii) (not shown here) yield very similar trends. Figure~\ref{fig:metrics_ns}(\textit{a}) shows that the kinetic energy content gradually drops with respect to the DNS as the radial resolution decreases, with a notable exception for the learned model $\tau \equiv \mathcal{M}$ which maintains a value within 5\% of the reference DNS. Figure~\ref{fig:metrics_ns}(\textit{b}) indicates that a decrease of the radial resolution goes along with an over-estimate of enstrophy for the under-resolved simulation $\tau \equiv 0$, opposite to the Leith model $\tau \equiv \nu_\text{L}$ which instead exhibits decreasing values of $\enstrophy$ on decreasing $N_s$. Only the hyperdiffusivity $\tau \equiv d(m)$ and learned models $\tau \equiv \mathcal{M}$ remain close to the reference DNS. Based on these observations, we decide to maintain the radial resolution $N_s$ for alternative models, i.e., hyperdiffusivity and Leith eddy-viscosity, similarly to \citet{chen2005large} and \citet{davy2026effect}. In other words, $N_s = 321$ is retained for the alternative models as the coarsest radial resolution at which their integrated diagnostics remain within DNS statistical uncertainty (see Fig.~\ref{fig:metrics_ns}); at the coarser resolution $N_s = 65$ used by the learned model $\tau \equiv \mathcal{M}$, both alternative models depart from the DNS by $50$--$80\%$ on $E_T$ and $\enstrophy$.
\begin{figure}
  \includegraphics[width=.95\linewidth]{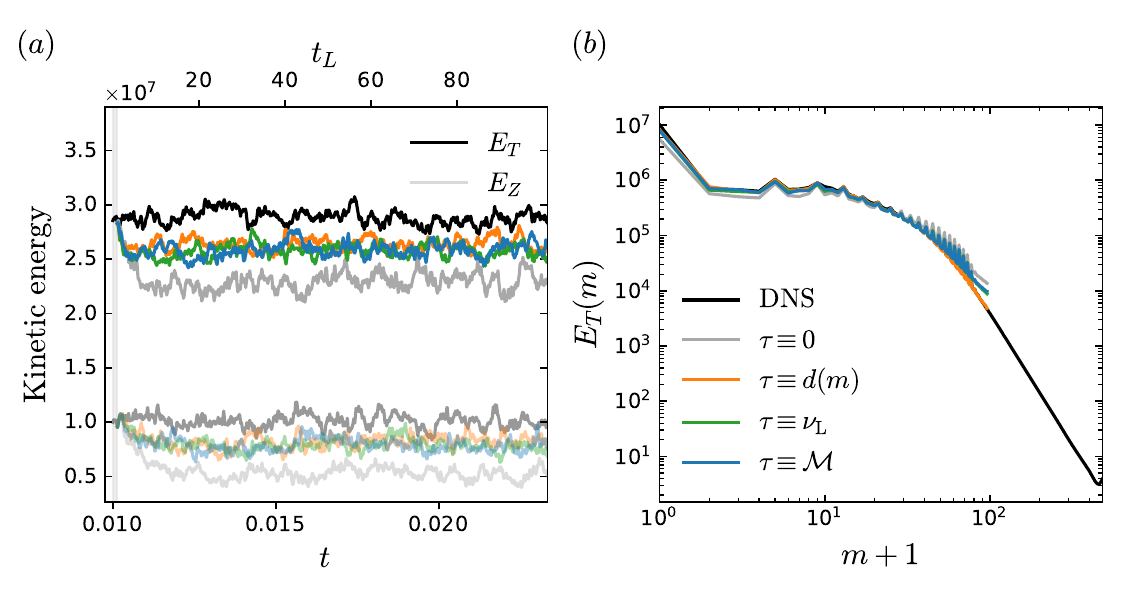}
  \caption{
  (\textit{a}) 
  Evolution of the total $E_{T}$ and zonal $E_{Z}$ kinetic energy densities as a function of time for configuration (iii). (\textit{b}) Time-averaged energy spectra as a function of the azimuthal wavenumber $m$ for configuration (iii). The highlighted region at the beginning of the kinetic energy evolution corresponds to the training dataset time span, which is equivalent to one turnover time. Simulations with the different models were started at this point and integrated for $100$ turnover time $\turnover$. \label{fig:kinetic-energy}}
\end{figure}

The impact of a chosen model on the energetic properties of a solution is illustrated in Figure~\ref{fig:kinetic-energy} for configuration (iii). Figure~\ref{fig:kinetic-energy}(\textit{a}) shows the time series of $E_T$ and $E_Z$ for the DNS, the different models $\tau \equiv \mathcal{M}$, $\tau \equiv \nu_\text{L}$, $\tau \equiv d(m)$, and the under-resolved simulation $\tau \equiv 0$. Once the steady state has been reached, the energy contained in the zonal jets corresponds to about 30\% of the kinetic energy content. 
We observe that the under-resolved simulation $\tau \equiv 0$ loses up to 30\% of the kinetic energy compared to the DNS in a few turnover times~$\turnover$. 
This loss can be mostly ascribed to the loss of mean zonal energy. Figure~\ref{fig:kinetic-energy}(\textit{b}) shows the corresponding time-averaged kinetic energy spectra as a function of azimuthal order $m$. The under-resolved simulation with $\tau \equiv 0$ presents a slight accumulation of kinetic energy at the tail of the spectrum. This accumulation is not observed with the hyperdiffusivity model $\tau \equiv d(m)$, which was precisely designed to prevent such pile-up. The under-resolved simulation $\tau \equiv 0$ also underestimates the zonal energy $(m = 0)$ and the energy contained in the intermediate scales $m \sim 1 - 30$. In contrast, the different models are able to match the kinetic energy spectra of the DNS up to $\overline{N_m}$, in spite of underestimating the total and zonal energy content by about 5--15\%, as can be seen in the time series of $E_T$ and $E_Z$ in Figure~\ref{fig:kinetic-energy}(\textit{a}).

\subsection{Integrated quantities}
For each configuration, Table~\ref{tab:evaluation-models} shows the time average of the Reynolds number, zonal  to total energy ratio, enstrophy, dissipation lengthscale, injected power, and the fraction of dissipation due to Ekman friction. Time averaging $\langle \cdot \rangle_t$ is performed over 100 turnover times, for the direct, $\tau \equiv \mathcal{M}$, $\tau \equiv \nu_\text{L}$, $\tau \equiv d(m)$, and $\tau \equiv 0$ numerical simulations.
\begin{table}
    \begin{center}
    \scalebox{0.9}{
    \begin{tabular}{l r c c c c c c c r}
    & \multicolumn{9}{r}{(i)} \\
    \hline
    & $(N_s, N_m)$ & $\langle Re \rangle_t$ & $\langle E_{Z} / E_{T} \rangle_t$ & $\langle \enstrophy \rangle_t$ & $\langle \dissls \rangle_t$ & $\langle \power \rangle_t$ & $\rfric$ & $\delta t_\text{eval}$ & $T_\text{eval}$ \\
    DNS & $(321, 400)$ & $6104$ & $0.265$ & $4.12 \times 10^{10}$ & $2.13 \times 10^{-2}$ & $2.51 \times 10^{11}$ & $0.671$ & $6 \times 10^{-8}$ & 1:41:11 \\
    $\tau \equiv 0$ & $(321, 80)$ & $5978$ & $0.178$ & $5.29 \times 10^{10}$ & $1.84 \times 10^{-2}$ & $2.67 \times 10^{11}$ & $0.603$ & $7 \times 10^{-8}$ & 57:19 \\
    $\tau \equiv d(m)$ & $(321, 80)$ & $5890$ & $0.231$ & $3.83 \times 10^{10}$ & $2.13 \times 10^{-2}$ & $2.57 \times 10^{11}$ & $0.702$ & $7 \times 10^{-8}$ & 57:08 \\
    $\tau \equiv \nu_\text{L}$ & $(321, 80)$ & $5828$ & $0.188$ & $3.99 \times 10^{10}$ & $2.06 \times 10^{-2}$ & $2.61 \times 10^{11}$ & $0.695$ & $8 \times 10^{-8}$ & 51:25 \\
    $\tau \equiv \mathcal{M}$ & $(65, 80)$ & $5836$ & $0.242$ & $3.79 \times 10^{10}$ & $2.12 \times 10^{-2}$ & $2.53 \times 10^{11}$ & $0.700$ & $5 \times 10^{-7}$ & \textbf{08:41} \\
    \hline
    & \multicolumn{9}{r}{(ii)} \\
    \hline
    & $(N_s, N_m)$ & $\langle Re \rangle_t$ & $\langle E_{Z} / E_{T} \rangle_t$ & $\langle \enstrophy \rangle_t$ & $\langle \dissls \rangle_t$ & $\langle \power \rangle_t$ & $\rfric$ & $\delta t_\text{eval}$ & $T_\text{eval}$ \\
    DNS & $(321, 480)$ & $6489$ & $0.240$ & $5.03 \times 10^{10}$ & $2.05 \times 10^{-2}$ & $1.60 \times 10^{11}$ & $0.373$ & $4 \times 10^{-8}$ & 2:36:33 \\
    $\tau \equiv 0$ & $(321, 96)$ & $5880$ & $0.182$ & $6.61 \times 10^{10}$ & $1.63 \times 10^{-2}$ & $1.80 \times 10^{11}$ & $0.264$ & $5 \times 10^{-8}$ & 1:17:46 \\
    $\tau \equiv d(m)$ & $(321, 96)$ & $6196$ & $0.218$ & $4.34 \times 10^{10}$ & $2.11 \times 10^{-2}$ & $1.61 \times 10^{11}$ & $0.461$ & $6 \times 10^{-8}$ & 1:05:14 \\
    $\tau \equiv \nu_\text{L}$ & $(321, 96)$ & $6242$ & $0.223$ & $4.71 \times 10^{10}$ & $2.04 \times 10^{-2}$ & $1.63 \times 10^{11}$ & $0.424$ & $6 \times 10^{-8}$ & 1:06:52 \\
    $\tau \equiv \mathcal{M}$ & $(65, 96)$ & $6260$ & $0.223$ & $4.65 \times 10^{10}$ & $2.06 \times 10^{-2}$ & $1.63 \times 10^{11}$ & $0.430$ & $3 \times 10^{-7}$ & \textbf{14:20} \\
    \hline
    & \multicolumn{9}{r}{(iii)} \\
    \hline
    & $(N_s, N_m)$ & $\langle Re \rangle_t$ & $\langle E_{Z} / E_{T} \rangle_t$ & $\langle \enstrophy \rangle_t$ & $\langle \dissls \rangle_t$ & $\langle \power \rangle_t$ & $\rfric$ & $\delta t_\text{eval}$ & $T_\text{eval}$ \\
    DNS & $(321, 480)$ & $7592$ & $0.348$ & $4.15 \times 10^{10}$ & $2.64 \times 10^{-2}$ & $1.63 \times 10^{11}$ & $0.489$ & $5 \times 10^{-8}$ & 1:47:37 \\
    $\tau \equiv 0$ & $(321, 96)$ & $6786$ & $0.238$ & $6.90 \times 10^{10}$ & $1.84 \times 10^{-2}$ & $1.97 \times 10^{11}$ & $0.300$ & $6 \times 10^{-8}$ & 55:38 \\
    $\tau \equiv d(m)$ & $(321, 96)$ & $7262$ & $0.313$ & $3.67 \times 10^{10}$ & $2.68 \times 10^{-2}$ & $1.72 \times 10^{11}$ & $0.573$ & $6 \times 10^{-8}$ & 55:30 \\
    $\tau \equiv \nu_\text{L}$ & $(321, 96)$ & $7177$ & $0.309$ & $4.03 \times 10^{10}$ & $2.53 \times 10^{-2}$ & $1.73 \times 10^{11}$ & $0.535$ & $8 \times 10^{-8}$ & 42:53 \\
    $\tau \equiv \mathcal{M}$ & $(65, 96)$ & $7192$ & $0.307$ & $4.03 \times 10^{10}$ & $2.54 \times 10^{-2}$ & $1.67 \times 10^{11}$ & $0.517$ & $4 \times 10^{-7}$ & \textbf{09:15} \\
    \end{tabular}
    }
    \caption{Statistical evaluation of the proposed models at resolution $(N_s, N_m)$ for each configuration explored in this study. Execution time has been measured in $T_\text{eval}$ for a single NVIDIA A-100 GPU with constant time step $\delta t_\text{eval}$, determined independently for each model and resolution as the largest empirically stable value found by trial and error. Reynolds number $Re$, zonal energy ratio $E_{Z} / E_{T}$, enstrophy $\enstrophy$, dissipation lengthscale $\dissls$, injected power $\power$ and ratio of dissipative processes $\rfric$ are time-averaged (denoted $\langle \cdot \rangle_t$) over $100$ turnover times.}
    \label{tab:evaluation-models}
    \end{center}
\end{table}
We observe a $\sim 5\%$ drop of the Reynolds number $\langle Re \rangle_t$ for all configurations with respect to the DNS, except for the under-resolved simulation $\tau \equiv 0$ that remains $10\%$ below the reference value.
The zonal energy ratio $E_{Z} / E_{T}$ is also not well reproduced by the under-resolved simulation which indicates that most of the zonal energy is lost, likely because it does not properly account for the subgrid transfers involved in the production of the axisymmetric velocity $\langle u_{\varphi} \rangle_{\varphi}$. The accumulation of energy at small scales is well reflected in the total enstrophy $\enstrophy$, which is always larger in the under-resolved simulations compared to their DNS counterparts. The simulations with hyperdiffusivity $\tau \equiv d(m)$ perform better on this metric, since they are effectively dissipating energy at small-scale (recall Fig.~\ref{fig:kinetic-energy} \textit{b}), but they suffer from a slightly excessive dissipation in the spherical container. Overall, the $\tau \equiv \mathcal{M}$, $\tau \equiv \nu_\text{L}$ and $\tau \equiv d(m)$ models are still able to successfully match the dissipation lengthscales of the DNS, which indicates a correct balance of energy and enstrophy production.
Regarding the fraction of energy dissipated through Ekman friction, $\rfric$, let us stress again that it is exactly computed via the balance \eqref{eq:tavg_eqbalance}, since the bulk viscous dissipation and injected power are directly measured. \citet{lemasquerier2023zonal} could not precisely measure the latter in their experiments; they instead produced estimates of the input power based either on the spectral distribution of the residual kinetic energy or on the assumption that energy was entirely dissipated by Ekman friction, $\rfric = 1$.
The relationship between these two estimates showed deviation from pure proportionality (see their Fig.~8(\textit{a})), which suggests that $\rfric$ was markedly below unity for some experiments. Here we find that it is about 67\% for configuration (i), 37\% for configuration (ii), and 49\% for configuration (iii), see Table~\ref{tab:evaluation-models}. The friction ratio is in all cases reproduced within 10\% by the different models. The energy budget, and hence $\rfric$, is however significantly altered for the under-resolved simulations. For the spherical configurations (ii) and (iii), the different models dissipate slightly more through Ekman friction compared to the reference DNS, while the $\tau \equiv 0$ simulations dissipate over 70\% of their energy through bulk viscous dissipation. 

Although the integrated quantities obtained with the different models ---excluding the under-resolved simulations--- remain close to those of the reference DNS, it is important to recall that the learned model $\tau \equiv \mathcal{M}$ operates on a coarser radial grid. This alleviates the CFL constraint associated with the radial direction and allows for significantly larger time steps, which we found to be approximately five times larger than those used with the alternative models. As a result, the learned model achieves a $\sim 5$–$7 \times$ reduction in time-to-solution while maintaining comparable performance across the considered metrics.

\subsection{Zonal structures}
\begin{figure}
  \includegraphics[width=.95\linewidth]{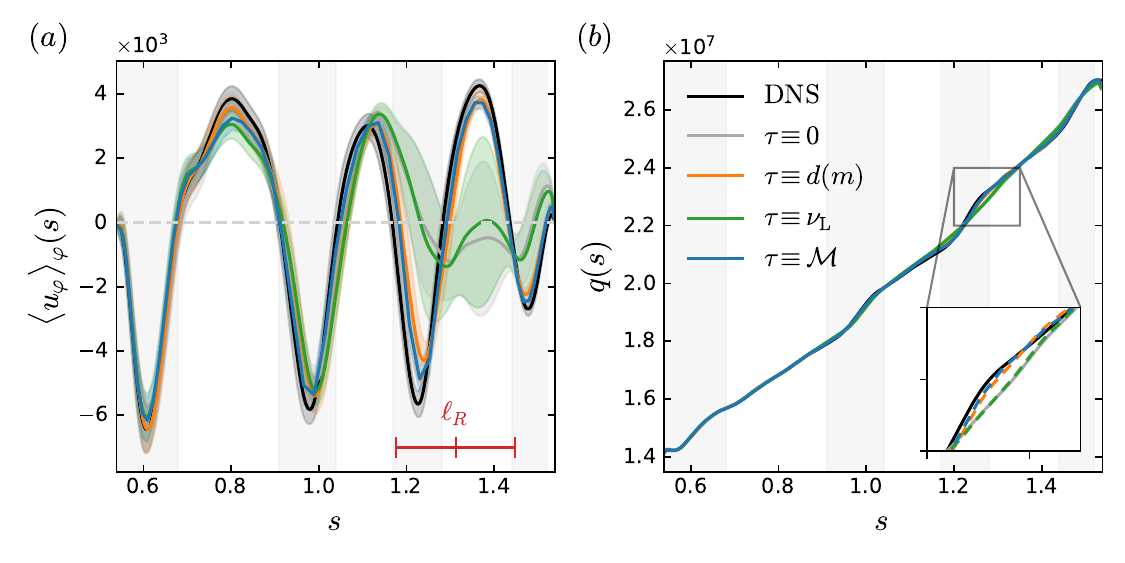}
  \caption{
  (\textit{a}) Time-averaged radial profile of the zonal velocity $\langle u_{\varphi} \rangle_\varphi$ and (\textit{b}) potential vorticity $q$ for configuration (i) which has constant $\beta$. 
  The alternating zonal jets on the outer boundary are almost entirely lost at coarse resolution $\tau \equiv 0$, but also with the Leith eddy-viscosity model $\tau \equiv \nu_\text{L}$. The horizontal segment in panel (\textit{a}) corresponds to the Rhines scale $\ell_R$  (Eq.~\ref{eq:rhines}) in this configuration. The inset in panel (\textit{b}) highlights the gradient of potential vorticity in the strongest retrograde jet; closure models are shown with dashed lines to improve readability. \label{fig:radial-profiles}}
\end{figure}

A hallmark of $\beta$-plane turbulence is the formation of zonal jets. The width of these jets is often found to be closely related to the so-called Rhines scale \citep{rhines1975waves}, here defined using the Rossby number, such that
\begin{equation}
    \ell_{R} = 2\pi \sqrt{\frac{Ro}{|\beta|}}.
    \label{eq:rhines}
\end{equation}
In the exponential container considered in configuration (i), $\beta$ is constant which makes the definition of $\ell_R$ well-posed. Figure~\ref{fig:radial-profiles}(\textit{a}) shows the time-averaged radial profiles of the zonal velocity $\langle u_\varphi \rangle_\varphi$ for this configuration. Seven jets of alternated directions of similar width are obtained. The segment in the bottom right corner indicates the good agreement between the jet widths and half the Rhines scale \citep[e.g.][]{heimpel2007turbulent}. The small standard deviations ---the shaded area--- highlight the stability of the jets over time in this configuration.
The different models accurately reproduce the jets pattern of the DNS, in stark contrast to the under-resolved simulation $\tau \equiv 0$ and the Leith eddy-viscosity $\tau \equiv \nu_\text{L}$ to a lesser extent, which fail to accurately capture the jet's structure for $s > 1.2$.

Zonal jets in quasi-geostrophic flows are directly related to the distribution of potential vorticity (PV) \citep{mcintyre2003potential} here defined by
\begin{equation}
    q = \frac{\langle \omega \rangle_\varphi + 2 E^{-1}}{h}.
\end{equation}
Regions of weak PV gradients hence correspond to prograde zonal jets, while retrograde jets coincide with regions of strong PV gradients. In the limit of quasi-geostrophic turbulence with $Re \gg 1$ and $Ro \ll 1$, the mixing of PV can yield the formation of staircases, i.e. successive regions of homogeneous PV separated by sharp interfaces 
\citep[e.g.][]{scott2012structure,verhoeven2014compressional}. 
Figure~\ref{fig:radial-profiles}(\textit{b}) shows that in our simulation the potential vorticity is almost linear with moderate steepening in the regions where the jets are retrograde. Given the moderate parameters considered here ---relatively large Ekman number and weak forcing---, the jets are still subjected to a strong role played by viscous friction and hence carry only a limited fraction of the total energy with $E_Z \approx 0.3 E_T$. It hence comes to no surprise that the level of staircasing remains mild \citep[see e.g.][for similar results at comparable parameters]{lemasquerier2023zonal}. 
This modest PV mixing is however sufficient to yield an asymmetry between the rounded prograde jets in regions of weaker PV gradient and the sharper retrograde jets when the gradient is stronger. This effect is well accounted for by the different models in regions where the jets structure has been preserved but mostly absent otherwise, as shown by the inset panel of Fig.~\ref{fig:radial-profiles}(\textit{b}).

\subsection{Spectral analysis}
In the presence of a $\beta$ effect, 2D rotating turbulent flows are expected to be anisotropic. In the so-called zonostrophic turbulence regime \citep[e.g.][]{sukoriansky2002universal,galperin2010geophysical}, the kinetic energy spectrum can be conveniently split into a residual component $E_R$ which accounts for non-axisymmetric motions and a zonal component $E_Z$, following 
\begin{equation}
    E_{R}(k) = C_R \epsilon^{2/3} k^{-5/3}, \quad
    E_{Z}(k_{s}) = C_{Z} \tilde{\beta}^{2} k_{s}^{-5},
    \label{eq:spec}
\end{equation}
where $\tilde{\beta} = 2\beta/E$, $k_s$ is the wavenumber in the direction orthogonal to the zonal flow and $k$ is the total wavenumber. The residual spectrum is expected to adhere to the classical Kolmogorov--Batchelor--Kraichnan (KBK) theory of 2D turbulence, where $C_R \approx 5-6$ is a universal constant and $\epsilon$ is the energy transfer rate \citep{boffetta2012two}. In the above equation, $C_{Z}$ is also a universal constant of order unity, in practice found to vary between $0.3$ and $3$ \citep[see e.g.][]{sukoriansky2007arrest,galperin2014cassini,cabanes2017laboratory,lemasquerier2023zonal}. 

To assess the relevance of the zonostrophic theory to our numerical simulations, we adopt a spectral transform appropriate to the cylindrical annulus, in a similar fashion to \citet{cabanes2024zonostrophic}. To do so, we decompose the velocity field components $(u_s, u_\varphi)$ using the so-called Weber--Orr transform, which can be expressed by \citep[see e.g.][]{wordsworth2008turbulence},
\begin{equation}
    u_{nm} = \int_{s_{i}}^{s_{o}} \int_{0}^{2 \pi} u(s, \varphi) \Psi_{nm}(s) e^{-\mathrm{ i } m \varphi} s \, \mathrm{d} s \, \mathrm{d} \varphi,
    \label{eq:weberorr}
\end{equation}
where $n$ and $m$ respectively denote the radial and azimuthal dimensionless wavenumbers. In the annulus, the support function $\Psi_{nm}$ is a linear combination of Bessel functions of the first and second kind $J_{m}$ and $Y_{m}$ such that \citep[e.g.][]{macrobert1932xvi}
\begin{equation}
    \Psi_{nm}(s) = Y_{m}(a_{nm} s_{o}) J_{m}(a_{nm} s) - J_{m}(a_{nm} s_{o}) Y_{m}(a_{nm} s).
\end{equation}
To enforce Dirichlet boundary conditions at both ends in the radial direction, the wavenumbers $a_{nm}$ have to fulfill the following transcendental equation \citep{cinelli1965extension}
\begin{equation}
    Y_{m}(a_{nm} s_{i}) J_{m}(a_{nm} s_{o}) - J_{m}(a_{nm} s_{i}) Y_{m}(a_{nm} s_{o}) = 0.
\end{equation}
From Eq.~\eqref{eq:weberorr}, we then compute the kinetic energy spectra using Parseval's identity
\begin{equation}
E_T = \sum_{n} \sum_{m=-N_m}^{N_m} E_{nm},
\end{equation}
where
\begin{equation}
    E_{nm} = \frac{1}{2}\dfrac{\pi^2}{s_o^2-s_i^2}\frac{a_{nm}^{2} J_{m}^{2}(a_{nm} s_{o})}{J_{m}^{2}(a_{nm} s_{i}) - Y_{m}^{2}(a_{nm} s_{o})} \left( \left| u_{s_{nm}} \right|^{2} + \left| 
    u_{\varphi_{nm}} \right|^2 \right)\,.
\end{equation}
Since $a_{nm}$ depends on the azimuthal order $m$, the wavenumbers $k$ are defined using the zonal wavenumber $k_n \equiv a_{n0}$ \citep{lemasquerier2023zonal}.
\begin{figure}
  \includegraphics[width=.95\linewidth]{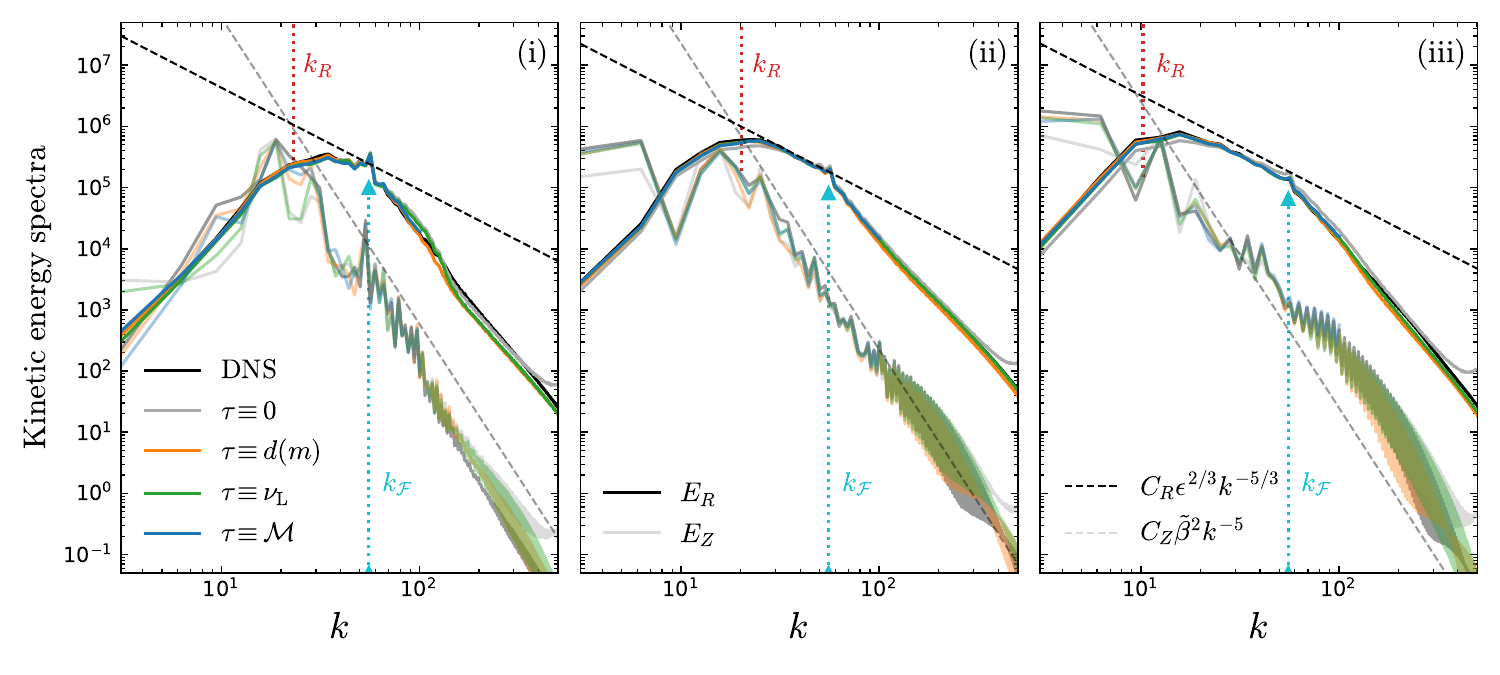}
  \caption{
  Time-averaged kinetic energy spectra of the residual ($E_{R}$) and zonal ($E_{Z}$) flow for each model and the three configurations studied. 
  The dashed lines correspond to the theoretical residual and zonal spectra assuming $C_R=5$ and $C_Z=(0.14,0.08,0.10)$ for the three considered cases. Vertical dotted lines mark the forcing and Rhines scales, $k_\mathcal{F}$ and $k_R$, in cyan and red, respectively. 
  For the spherical cases (ii) and (iii), $\tilde{\beta}$ is evaluated at mid-depth. \label{fig:bessel-fourier-spectra}}
\end{figure}
The spectra of the non-axisymmetric residuals with $m \neq 0$ are then constructed by summation of the 2D spectral contribution $E_{nm}$ into the bins defined by $a_{n0}$. To that end, we introduce $\mathcal{R}(p)$ as the ensemble of wavenumber indices $(n,m)$ which satisfy $a_{n0} - \delta k_n / 2 \leq a_{nm} \leq a_{n0} + \delta k_n / 2$, where $\delta k_n = k_{n+1}-k_n \approx \pi $ is the almost equidistant bin spacing. Energy conservation then demands that $\sum_p E(k_p) \delta k_p = \sum_n \sum_{m=-N_m}^{N_m} E_{nm}$ such that $E(k_p) = 1/\delta k_p \sum_{(n,m) \in \mathcal{R}(p)} E_{nm}$. The dimension of $E(k_p)$ in the above expressions is a cubed length scale per squared time unit.

Figure~\ref{fig:bessel-fourier-spectra} shows the time-averaged zonal spectra $E_{Z}(k)$ and residual spectra $E_{R}(k)$ respectively computed over $5000$ and $100$ statistically-independent snapshots. 
For each configuration, vorticity is injected at wavenumber $k_\mathcal{F}$ which is inversely proportional to the vortex spacing $\Delta_\mathcal{F}$ with $k_{\mathcal{F}} = \pi \sqrt{2} / \Delta_{\mathcal{F}} \approx 55$ (dotted cyan lines). This corresponds to the localized peaks of residual energy visible in the three configurations for $k\approx 50-60$. 
The forcing hence mostly occurs at a much larger scale than the truncation of the coarse-grained models, to ensure that the model correction only includes energy transfers related to subgrid processes. The energy transfer rate can be estimated by $\epsilon=\power$ once the statistically-steady state has been reached.

For the three considered configurations, we observe a limited range of scales ---slightly broader in configurations (ii) and (iii) compared to (i)--- over which the residual spectra adhere to the theoretical scaling $C_R\epsilon^{2/3} k^{-5/3}$ (dashed black lines), taking $C_R=5$. At scales smaller than the forcing scale, the residual spectra drop with a slope close to~$-5$, steeper than the expected~$-3$ scaling due to the direct cascade of enstrophy in 2D turbulence \citep{boffetta2012two}. Such steeper scaling behaviors were also reported by \citet{lemasquerier2023zonal} and are attributed to the role played by the large-scale drag \citep{boffetta2007energy}. While the different models accurately captures the residual spectra up to the scale of the numerical truncation, the under-resolved simulations slightly lack energetic content at large scale while accumulating energy at small scale.
Regarding the zonal spectra, we recover in configurations (i) and (ii) a range of scales where the spectra also conform to the theoretical scaling $C_Z\tilde{\beta}^2 k^{-5}$ with $C_Z \approx 0.1$ (dashed gray lines).
The $-5$ slope is less obvious in configuration (iii) and limited to a narrower range of wavenumbers. One possible explanation for this difference is that in configuration (ii) the jets are mostly located in the lower half of the domain, whereas in configuration (iii) they are developed throughout the entire fluid volume (recall Fig.~\ref{fig:configurations-fields}). 
This implies that the variability of $\beta(s)$ in the spherical container will have a greater impact on the zonal spectrum in configuration (iii) than in configuration (ii).
The value obtained for $C_Z$ is in the low range of published values, consistent with those obtained by \citet{lemasquerier2023zonal} for a configuration with a similar topographic $\beta$ effect. The agreement with the $-5$ slope approximately stops at the Rhines scale $k_R = 2\pi/\ell_R$ (dotted red lines), which is also close to the most energetic scale in the three configurations.
We confirm that our configurations are not in a zonostrophic regime by observing a zonostrophy index $R_\beta \approx 1$. This index is defined as $R_\beta = k_\beta / k_R$, where $k_\beta$ is the transitional wavenumber which marks the crossing between the residual and the zonal spectra. While the $-5$ slope is observed in each model, the zonal spectra of the under-resolved simulations show some significant differences, in particular at $k < k_\mathcal{F}$. In contrast, the different models correctly reproduce the entire zonal spectrum.

\subsection{Temporal patterns: Waves and jets}
\begin{figure}
  \includegraphics[width=.9\linewidth]{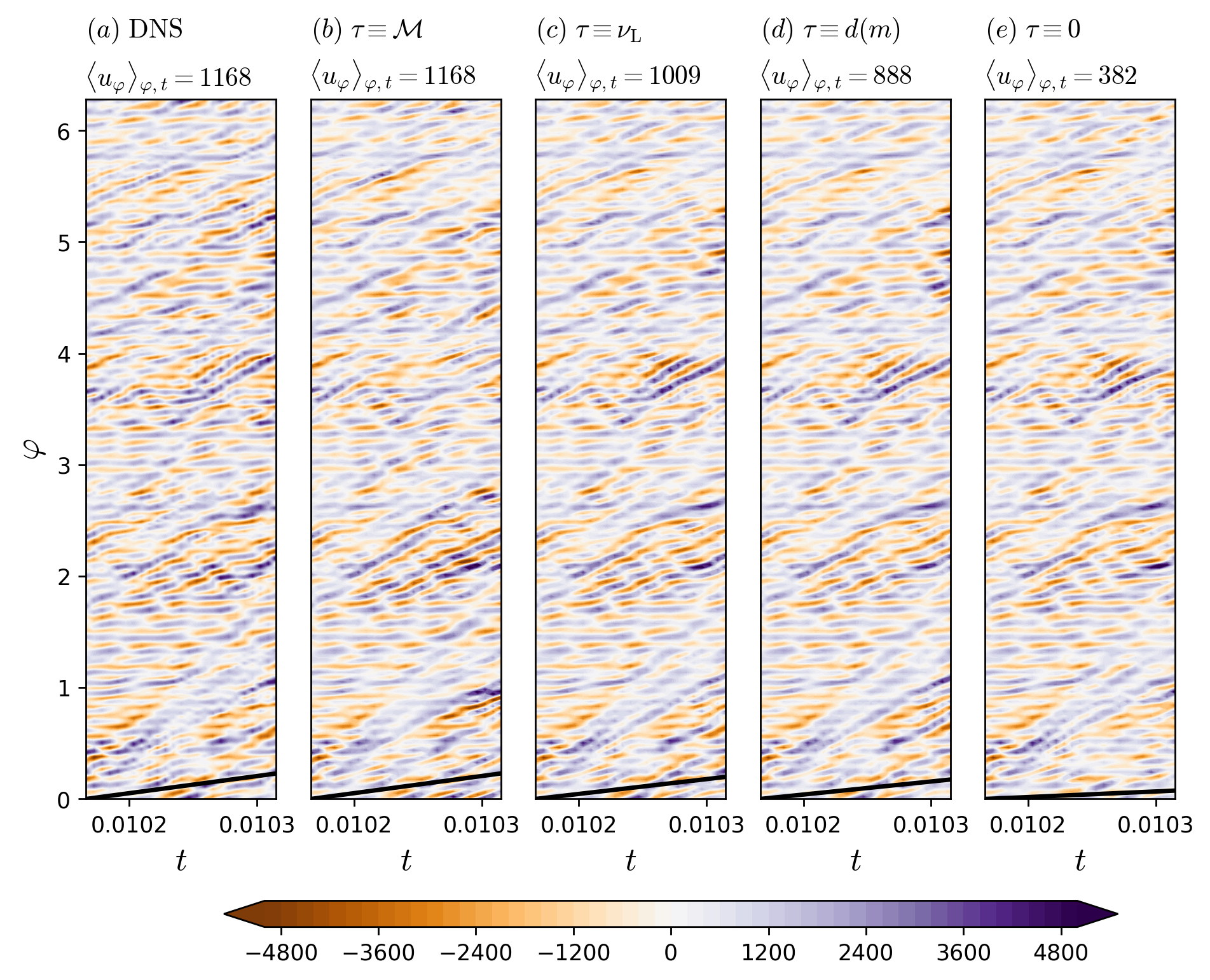}
  \caption{(\textit{a}) Hovmöller diagrams (time-azimuth) of the radial velocity $u_{s}$ at a fixed radius $s = 1.32$ inside the Rossby waves region for configuration (ii) computed with the DNS. (\textit{b-d}) Same quantity computed using the different models and (\textit{e}) for the under-resolved simulation.  
  The time interval approximately corresponds to one single turnover time. The solid black line shows the drift due to the advection by zonal velocity, with $\langle u_\varphi \rangle_{\varphi,t}(s=1.32)$ given for each simulation at the top of their respective panel. \label{fig:rossby-waves}}
\end{figure}

Finally, we explore the temporal dependence of Rossby waves and large-scale jets in the vicinity of the outer boundary for configuration (ii).  
Close to the outer boundary, as seen previously in Fig.~\ref{fig:configurations-fields}(\textit{c}), Rossby waves prevail. 
Figure~\ref{fig:rossby-waves} shows a longitude-time map of the cylindrically radial velocity $u_s$ at $s=1.32$ over a single turnover time. We observe the evolution of non-axisymmetric patterns which results from the superposition of two effects: advection by the average background zonal flow at this radius, indicated by the black line in Fig.~\ref{fig:rossby-waves}, and additional prograde transport by Rossby waves, since patterns appear steeper than would be expected from pure eastward advection by the local zonal flow.
The different models and the under-resolved simulation successfully predict these rapidly evolving small-scale structures. We find however that advection by the zonal flow is weaker for the under-resolved simulation in Fig.~\ref{fig:rossby-waves}(\textit{e}) ---in comparison with the reference DNS---, but still preserves the quasi-linear conditions necessary for Rossby wave propagation in this region.

\begin{figure}
  \includegraphics[width=.95\linewidth]{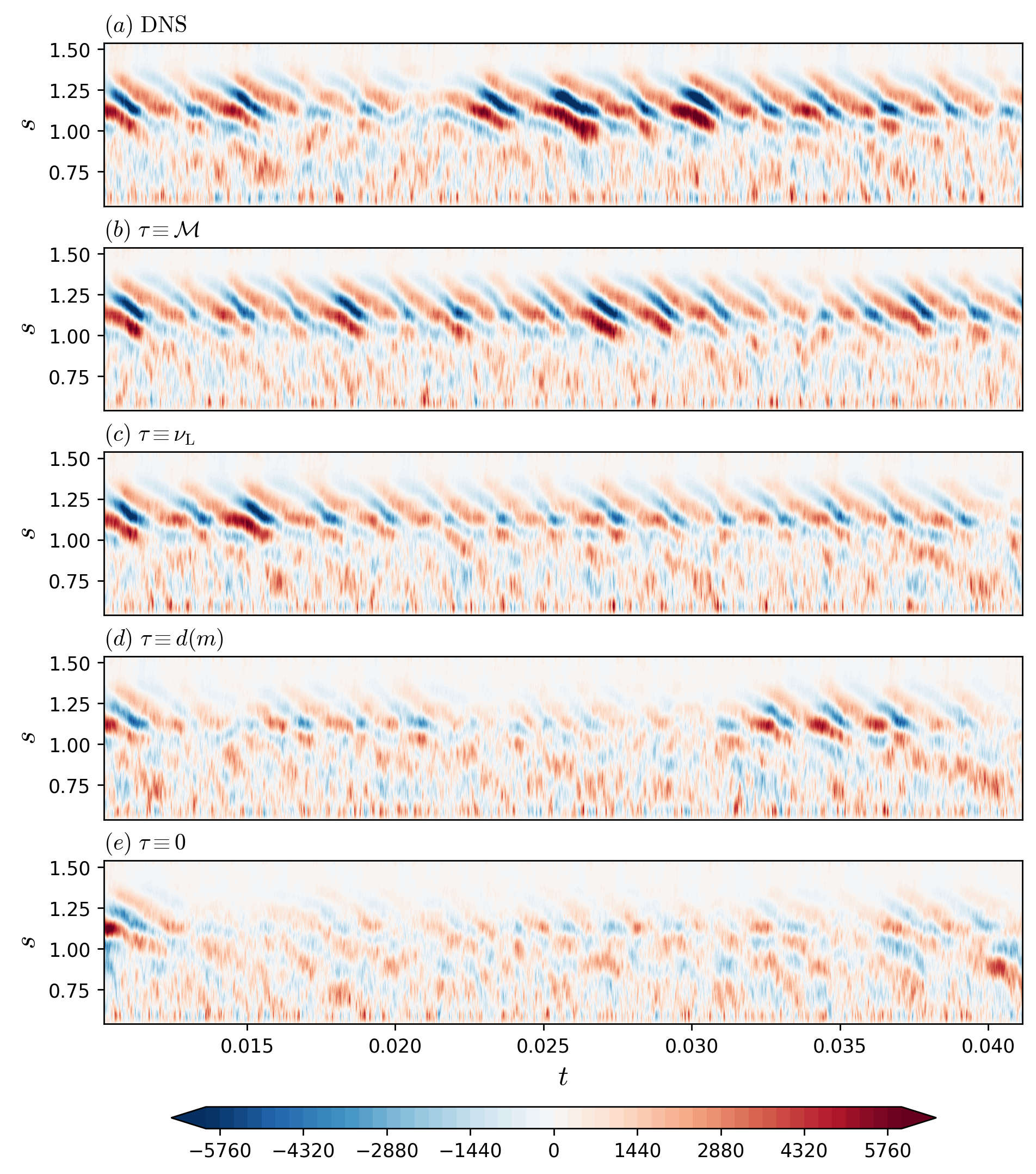}
  \caption{(\textit{a}) Time evolution of the zonal velocity with respect to its time-averaged profile $\langle u_\varphi \rangle_\varphi(s,t) - \langle u_\varphi \rangle_{\varphi,t}(s)$ for configuration (ii) computed with the DNS. (\textit{b-d}) Same quantity computed using the different models and (\textit{e}) for the under-resolved simulation. The time interval corresponds to $200$ turnover times. \label{fig:jets-drift}}
\end{figure}

Figure~\ref{fig:jets-drift} shows the time-evolution of the mean zonal velocity with respect to its time-averaged value over a longer temporal horizon of $200$ turnover times $\turnover$. 
Figure \ref{fig:jets-drift}(\textit{a}) reveals that jets in the DNS exhibit a quasi-periodic inward drift for radii roughly spanning the range $s\in[1,1.4]$, with a recurrence time scale of about $10$~$\turnover$. This is in stark contrast with the dynamical behavior of the zonal jets in the exponential container (case i), which do not migrate over time (see Fig.~\ref{fig:radial-profiles}(\textit{a}) for their radial time-averaged profiles).
In the case of thermal convection in a rapidly rotating thick spherical shell, \citet{rotvig2007multiple} reported a similar drifting pattern of zonal jets using both the three-dimensional description of the system and its quasi-geostrophic approximation \citep[see also][]{guervilly2017multiple}. \citet{lonner2022planetary} evidenced zonal jets which rather migrate outwards in their laboratory experiments of rotating convection. The jets' drift direction differs from the spherical case due to $\beta$ carrying the opposite sign in their experiments which involve a free paraboloidal surface. One possible explanation for the jet migration, put forward by \citet{chemke2015poleward}, is related to the variations of $\beta(s)$, which breaks the symmetry of the eddy momentum flux convergence $\langle u_s \omega \rangle_\varphi$ between the inward and the outward flank of a single jet \citep[see also][Fig.~5.15]{cope2021dynamics}. \citet{lemasquerier2021zonal} have shown that Reynolds stress amplification could indeed occur at critical radii where the Rossby waves are stationary, i.e. when their phase speed in absence of zonal flow is opposite to the jet velocity. In a spherical container as considered in case (ii), $\beta$ is negative and the phase speed of Rossby waves is positive; such an amplification is then only susceptible to occur in retrograde jets with $\langle u_\phi \rangle_\varphi(s) < 0$. Since $\beta(s)$ also decreases with radius, this condition will be met on the inward flank of retrograde jets, explaining the direction of migration visible in Fig.~\ref{fig:jets-drift}.
The jets drift is almost lost for the hyperdiffusivity model and under-resolved simulation (see Fig.~\ref{fig:jets-drift} \textit{d,e}).
The only models capable of reproducing those drifts at a comparable intensity are $\tau \equiv \mathcal{M}$ and $\tau \equiv \nu_\text{L}$, as illustrated by Fig.~\ref{fig:jets-drift}(\textit{b}) and Fig.~\ref{fig:jets-drift}(\textit{c}). 

\section{Discussion \label{sec:discussion}}
We studied two-dimensional quasi-geostrophic turbulence driven by a stationary array of Gaussian vorticity sources in an annulus with no-slip boundary conditions, considering three configurations that differ in container geometry and rotation rate. The flow results from the interplay of 2D turbulence, Rossby wave propagation due to the $\beta$-effect ---the variation of the effective planetary vorticity with the distance to the rotation axis--- and large-scale Ekman friction. In the exponential container, the spatially constant $\beta$-effect produces homogeneous turbulence with three pairs of alternating zonal jets of comparable size. Both spherical configurations exhibit a marked spatial segregation: a turbulence-dominated inner region coexists with an outer layer governed by Rossby-wave dynamics, reflecting the radially decreasing $\beta(s)$ in these geometries.

A spectral analysis using cylindrical harmonic functions shows that, although the fraction of energy carried by zonal jets remains modest, all three flows conform to the predictions of zonostrophic turbulence theory \citep[e.g.][]{sukoriansky2002universal}, consistent with the numerical and laboratory results in \cite{lemasquerier2023zonal} and \cite{zhu2025zonal}. The temporal dynamics exhibit a separation of timescales: short-term variability is governed by Rossby wave propagation, while on longer timescales the zonal jets in the spherical container undergo a quasi-periodic inward drift.
This jet migration, documented in a variety of QG settings \citep{rotvig2007multiple,guervilly2017multiple,chemke2015poleward}, appears here to be directly controlled by the spatial variation of $\beta$ with cylindrical radius, following the mechanism proposed by \citet{chemke2015poleward}. The physical ingredients governing the migration velocity and the quasi-periodicity of the phenomenon, however, remain to be clarified \citep{cope2021dynamics}.

We proposed and assessed a subgrid-scale (SGS) model correction for this fluid mechanical system resting on neural networks. 
The bounded geometry introduces specific technical challenges for SGS modeling. The spectral Chebyshev collocation method requires a Galerkin projection to generate filtered datasets while preserving boundary conditions. 
More critically, the non-homogeneity of the radial direction yields a commutation error in the definition of the SGS term, which makes ``offline'' learning impractical: in practice, models trained offline lead to unstable simulations.  ``Online'' learning sidesteps this difficulty entirely, since it operates over the coarse states without explicitly evaluating the SGS terms.  
The training timespan was set to one eddy turnover time, estimated from the vorticity decorrelation time yielding datasets of fewer than $1000$ snapshots.

Across all three configurations, the online-trained model is stable and accurate over integrations at least 100 times longer than the training window. It reproduces kinetic energy budgets, spectral energy distributions, and the spatial structure of the zonal jets.  
The model also captures the slow quasi-periodic jet drift in the spherical configurations ---a phenomenon operating on timescales an order of magnitude larger than the training window--- demonstrating that the learned correction includes the relevant subgrid physics. 

We also assessed the performance of classical closure models, namely hyperdiffusivity and the Leith model for 2D turbulence \citep{leith1996stochastic}. Provided filtering is restricted to the periodic azimuthal direction, both account well for integrated diagnostics, energy spectra, and short-term Rossby-wave dynamics. 
However, a further reduction in the bounded radial resolution significantly degrades their accuracy. The neural-network model, by contrast, enables such a reduction and thereby delivers a concrete computational gain: on a single NVIDIA A-100 GPU, integrating over $100$ turnover times for configuration~(i) takes $8.6$~minutes with the learned model, compared to more than $45$~minutes for the classical closures and $1.6$~hours for the DNS, an $11$-fold reduction relative to the DNS and a $5$-fold reduction relative to the classical models. Since hyperdiffusivity has a negligible impact on the time-to-solution and adequately handles azimuthal subgrid transfers, a spatially-coupled strategy combining hyperdiffusivity in the azimuthal direction with a neural-network correction restricted to the
radial direction could further reduce model complexity and cost. 

These results open new perspectives for systematically exploring rare events occurring on long timescales, such as jet drift and  merging in $\beta$-plane turbulence \citep[e.g.][]{simonnet2021multistability}. More broadly, the framework retains the physical structure of the underlying numerical solver: only the subgrid closure is learned, while the forcing, Coriolis, friction and viscous operators keep their usual discretized form, and the associated diagnostics, such as the power balance underlying $\rfric$ in Table~\ref{tab:evaluation-models} remains well defined for every model considered, including $\tau \equiv \mathcal{M}$. This does not mean that $\tau \equiv \mathcal{M}$ reproduces the DNS energy content exactly: like the other reduced models, it underestimates the total and zonal kinetic energy by $5$--$15\%$ relative to the DNS (see Fig.~\ref{fig:kinetic-energy}). The situation is different for purely data-driven approaches such as \citet{shokar2024stochastic}, which replace the governing equations themselves with a learned mapping: none of these operators is represented in such a model, so this kind of diagnostic cannot even be formulated for its output. The present approach also requires comparatively little training data, since the closure is the only learned component and the solver supplies the rest of the dynamics. This is particularly attractive for more turbulent regimes: the present configurations are limited to $Re \sim 10^4$ by the computational cost of the dense Chebyshev collocation matrices, but switching to a sparse implementation \citep[e.g.][]{marti2016computationally} would allow reaching $Re \sim 10^6$ while retaining the same numerical structure. In this context, training data could be generated independently using a distributed high-performance code such as \texttt{pizza} \citep{gastine2019pizza}, with the differentiable collocation solver only mobilised for the training and evaluation phases alone. A natural next step is to extend the methodology to convection-driven flows, which introduce a coupled evolution equation for the temperature perturbation requiring its own subgrid closure.

\begin{acknowledgments}
This work has been supported by the European High-Performance Computing Joint Undertaking (JU) under Grant Agreement No. 101093038 (ChEESE-2P). 
As part of the "France 2030" initiative, this work has benefited from a national grant managed by the French National Research Agency (Agence Nationale de la Recherche) attributed to the Exa-DI project of the NumPEx PEPR program, under the Reference No. ANR-22-EXNU-0006.
Numerical computations were performed on the S-CAPAD/DANTE platform, IPGP, France.
\end{acknowledgments}

\section*{Data Availability Statement}
The code used to generate data, train and evaluate models has been made openly available as a collection of scripts and notebooks in \url{https://github.com/hrkz/online-sgs-qg-planetary} and archived in \cite{hugo_frezat_2026_22915270} under the MIT licence.

\appendix

\section{Ekman pumping contribution \label{app:ekman-exponential}}
In this appendix, we recall the Ekman pumping contribution outside the tangent cylinder in a spherical container and compute the contribution in an exponential container, in cylindrical coordinates. The Ekman pumping flow is given by \citet{greenspan1968theory},
\begin{align}
    \bm{u} \bcdot \bm{n} |_{\pm h} &= -\frac{E^{1/2}}{2} \bm{n} \bcdot \bm{\nabla} \times \left( \frac{\bm{n} \times \bm{u} \pm \bm{u}}{\sqrt{| \bm{n} \bcdot \bm{e}_{z} |}} \right) \nonumber \\
    &= \bm{n} \bcdot \left[ \frac{\bm{\nabla} \times (\bm{n} \times \bm{u} \pm \bm{u})}{\sqrt{| \bm{n} \bcdot \bm{e}_{z} |}} + \bm{\nabla} \left( \frac{1}{\sqrt{| \bm{n} \bcdot \bm{e}_{z} |}} \right) \times (\bm{n} \times \bm{u} \pm \bm{u}) \right] \label{eq:ekman-pumping}
\end{align}
where $\bm{n}$ is the outward unit vector normal to the surface. From now on, we assume that the vertical velocity $u_z$ increases linearly with $z$ such that
\begin{equation}
    \frac{\partial u_{z}}{\partial z} = \beta u_{z} = \frac{1}{h} \frac{\mathrm{d} h}{\mathrm{d s}} u_{s}
\end{equation}
In a spherical container with $h = (s_{o}^{2} - s^{2})^{1/2}$, Ekman pumping is well known, e.g. \citep{schaeffer2005quasigeostrophic} and can be expressed as,
\begin{equation}
    \bm{u} \bcdot \bm{n}|_{\pm h} = \frac{s}{s_{o}} u_{s} \pm \frac{h}{s_{o}} u_{z} = -\frac{E^{1/2}}{2} \frac{h^{1/2}}{s_{o}^{1/2}} \left[ \omega_{z} + \frac{s}{2 h^{2}} u_{\varphi} + \frac{5}{2} \frac{s_{o}s}{h^{3}} u_{s} - \frac{s}{h^{2}} \frac{\partial u_{s}}{\partial \varphi} \right].
\end{equation}
As discussed in Sec. \ref{sec:governing_equations}, we only retain the term proportional to vorticity  in our implementation. The stretching term used in \eqref{eq:sph-qg-omega} and \eqref{eq:sph-qg-uphi-axisymmetric} for the spherical container then reads
\begin{equation}
    \frac{2}{E} \frac{\partial u_{z}}{\partial z} = - \frac{2}{E} \frac{s}{h^{2}} u_{s} + \underbrace{\frac{s_{o}^{1/2}}{E^{1/2} h^{3/2}}}_{\Upsilon} \omega_{z}.
\end{equation}
Let us now derive \eqref{eq:ekman-pumping} for an exponential container with $h = \exp(\beta s)$. Let us expand the first part using the following identity,
\begin{equation}
    \bm{\nabla} \times (\bm{n} \times \bm{u} \pm \bm{u}) = \bm{n} (\bm{\nabla} \bcdot \bm{u}) - \bm{u}(\bm{\nabla} \bcdot \bm{n}) + \bm{\nabla n} \bcdot \bm{u} - \bm{\nabla u} \bcdot \bm{n} \pm \bm{\omega}, \label{eq:first-part-identity}
\end{equation}
where $\bm{\omega} = \bm{\nabla} \times \bm{u}$ is the vorticity.
The outward unit normal to the exponential container surface is expressed by
\begin{equation}
    \bm{n} = \frac{1}{(1 + \beta^{2} h^{2})^{1/2}}
    \begin{pmatrix} 
    -\beta h & 0 & \pm 1
    \end{pmatrix}^{\mathsf{T}},
    \label{eq:normal-vector}
\end{equation}
where $+1$ (resp. $-1$) is to be used in the $z > 0$ (resp. $z < 0$) half-space. 
The corresponding gradient and divergence read
\[
    \bm{\nabla} \bm{n} = 
    \begin{pmatrix}
    \dfrac{\beta^{2} h}{(1 + \beta^{2} h^{2})^{3/2}} & 0 & 0 \\
    0 & -\dfrac{\beta h}{s (1 + \beta^{2} h^{2})^{1/2}} & 0 \\
    \pm\dfrac{\beta^{3} h^{2}}{(1 + \beta^{2} h^{2})^{3/2}} & 0 & 0
    \end{pmatrix},
\quad
    \bm{\nabla} \bcdot \bm{n} = -\frac{\beta^{2} h}{(1 + \beta^{2} h^{2})^{3/2}} - \frac{\beta h}{s (1 + \beta^{2} h^{2})^{1/2}}\,.
\]
In cylindrical coordinates, the gradient of the velocity $\bm{\nabla u}$ reads, assuming that $\partial u_{s} / \partial z = \partial u_{\varphi} /\partial z = 0$,
\begin{equation}
    \bm{\nabla u} = 
    \begin{pmatrix}
    \dfrac{\partial u_{s}}{\partial s} & \dfrac{1}{s} \dfrac{\partial u_{s}}{\partial u_{\varphi}} - \dfrac{u_{\varphi}}{s} & \dfrac{\partial u_{z}}{\partial z} \\
    \dfrac{\partial u_{\varphi}}{\partial s} & \dfrac{1}{s} \dfrac{\partial u_{\varphi}}{\partial \varphi} + \dfrac{u_{s}}{s} & \dfrac{\partial u_{\varphi}}{\partial z} \\
    \dfrac{\partial u_{z}}{\partial s} & \dfrac{1}{s} \dfrac{\partial u_{z}}{\partial \varphi} & \dfrac{\partial u_{z}}{\partial z}
    \end{pmatrix} = 
    \begin{pmatrix}
    \dfrac{\partial u_{s}}{\partial s} & \dfrac{1}{s} \dfrac{\partial u_{s}}{\partial u_{\varphi}} - \dfrac{u_{\varphi}}{s} & 0 \\
    \dfrac{\partial u_{\varphi}}{\partial s} & \dfrac{1}{s} \dfrac{\partial u_{\varphi}}{\partial \varphi} + \dfrac{u_{s}}{s} & 0 \\
    \dfrac{\partial u_{z}}{\partial s} & \dfrac{1}{s} \dfrac{\partial u_{z}}{\partial \varphi} & \pm \dfrac{1}{h} \dfrac{\mathrm{d} h}{\mathrm{d} s} u_{s}
    \end{pmatrix}. \nonumber
\end{equation}
Substituting these quantities in the identity \eqref{eq:first-part-identity}, omitting the calculation in $\varphi$ since $n_{\varphi} = 0$, we get the following
\begin{equation}
    \bm{\nabla} \times (\bm{n} \times \bm{u} \pm \bm{u}) = 
    \begin{pmatrix}
    \dfrac{\beta h}{s (1 + \beta^{2} h^{2})^{1/2}} u_{s} + \dfrac{\beta h}{(1 + \beta^{2} h^{2})^{1/2}} \dfrac{\partial u_{s}}{\partial s} + \dfrac{\beta h}{s} \dfrac{\partial u_{s}}{\partial \varphi} \\
    \cdots \\
    \pm \dfrac{\beta^{2} h^{2}}{s (1 + \beta^{2} h^{2})^{1/2}} u_{s} \pm \dfrac{\beta}{(1 + \beta^{2} h^{2})^{1/2}} u_{s} \pm \dfrac{\beta^{2} h^{2}}{(1 + \beta^{2} h^{2})^{1/2}} \dfrac{\partial u_{s}}{\partial s}\pm \omega_{z}
    \end{pmatrix}. \nonumber
\end{equation}
Then, the first part of the Ekman contribution \eqref{eq:ekman-pumping} simplifies to,
\begin{equation}
    \frac{1}{\sqrt{| \bm{n} \bcdot \bm{e}_{z} |}} \bm{n} \bcdot
    \bm{\nabla} \times (\bm{n} \times \bm{u} \pm \bm{u}) = \frac{1}{(1 + \beta^{2} h^{2})^{1/4}} \left[ \omega_{z} - \frac{\beta}{(1 + \beta^{2} h^{2})^{1/2}} u_{s} - \frac{\beta^{2} h^{2}}{s} \frac{\partial u_{s}}{\partial \varphi} \right]. \label{eq:ekman-first-part}
\end{equation}
For the second part of the Ekman contribution, let us first observe that relationship with planetary vorticity $f$,
\begin{equation}
    \bm{\nabla} \left( \frac{1}{\sqrt{| \bm{n} \bcdot \bm{e}_{z} |}} \right) = f \bm{e}_{s} = \frac{\beta^{3} h^{2}}{2 (1 + \beta^{2} h^{2})^{3/4}} \bm{e}_{s}, \label{eq:planetary-vorticity}
\end{equation}
and thus
\begin{equation}
    \bm{n} \bcdot f \bm{e}_{s} \times (\bm{n} \times \bm{u} \pm \bm{u}) = \frac{1}{\sqrt{1 + \beta^{2} h^{2}}} \left[ f \left( 1 + \beta^{2} h^{2} \right)^{1/2} u_{s} + f u_{\varphi} \right]. \nonumber
\end{equation}
Replacing $f$ from \eqref{eq:planetary-vorticity}, we obtain the second part of the Ekman contribution \eqref{eq:ekman-pumping},
\begin{equation}
    \bm{n} \bcdot \bm{\nabla} \left( \frac{1}{\sqrt{| \bm{n} \bcdot \bm{e}_{z} |}} \right) \times (\bm{n} \times \bm{u} \pm \bm{u}) = \frac{1}{(1 + \beta^{2} h^{2})^{1/4}} \left[ \frac{\beta^{3} h^{2}}{2 (1 + \beta^{2} h^{2})^{1/2}} u_{s} + \frac{\beta^{3} h^{2}}{2 (1 + \beta^{2} h^{2})} u_{\varphi} \right]. \label{eq:ekman-second-part}
\end{equation}
Finally, combining \eqref{eq:ekman-first-part} and \eqref{eq:ekman-second-part} gives,
\begin{equation}
    \bm{u} \bcdot \bm{n} |_{\pm h} = -\frac{E^{1/2}}{2} \frac{1}{(1 + \beta^{2} h^{2})^{1/4}} \left[ \omega_{z} - \frac{2 \beta - \beta^{3} h^{2}}{2 (1 + \beta^{2} h^{2})^{1/2}} u_{s} - \frac{\beta^{2} h^{2}}{s} \frac{\partial u_{s}}{\partial \varphi} + \frac{\beta^{3} h^{2}}{2 (1 + \beta^{2} h^{2})} u_{\varphi} \right]. \label{eq:ekman-pumping-exponential} 
\end{equation}
Now, observe that expanding $\bm{u} \bcdot \bm{n} |_{\pm h}$ using \eqref{eq:normal-vector} gives an expression for $u_{z}$ as a function of the Ekman pumping.
Using a difference formula $\partial u_{z} / \partial z = (u_{z}(h) - u_{z}(-h))/2h$, the stretching term used in \eqref{eq:sph-qg-omega} and \eqref{eq:sph-qg-uphi-axisymmetric} for the exponential container can be derived similarly than for the spherical container by substituting $u_{z}$ using \eqref{eq:ekman-pumping-exponential} and neglecting the non-dominant term,
\begin{equation}
    \frac{2}{E} \frac{\partial u_{z}}{\partial z} = \frac{2}{E} \beta u_{s} + \underbrace{\frac{(1 + \beta^{2} h^{2})^{1/4}}{E^{1/2} h}}_{\Upsilon} \omega_{z}.
\end{equation}

\section{Fourier--Chebyshev numerical discretization \label{app:numerical-discretisation}}
Since the boundary conditions \eqref{eq:boundary-conditions} are separable, any unknown field $f$ can be expanded upon the tensor product of a periodic basis in the azimuthal direction and a bounded basis in the radial direction.

The expansion for the azimuthal direction is carried out by truncated (real) Fourier series on equally-spaced grid points $\varphi_{k} \in [0, 2\pi)$,
\begin{equation}
    f(s, \varphi_k) \approx \sideset{}{'}\sum_{m = 0}^{N_{m}} \Re \left\{ f_{m}(s) \, e^{\mathrm{ i } m \varphi_{k}} \right\},
\end{equation}
where the prime summation indicates that $\forall m > 0$, $f_{m}$ coefficients are multiplied by two to account for the complex conjugate symmetry $f_{-m}^{*} = f_{m}$. The number of grid points in the azimuthal direction is set to $N_{\varphi} = 3N_{m}$ to prevent aliasing from the quadratic nonlinear terms \citep{boyd2001chebyshev,canuto2006spectral}.

In the radial direction, a Chebyshev collocation method is employed on a grid of $N_{s}$ points $s_k$; these points are the images of the Gauss--Lobatto points
\begin{equation}
    x_{k} = \cos \left[ \frac{\pi (k - 1)}{N_{s} - 1} \right], \quad 1 \leq k \leq N_{s},
    \label{eq:GL_grid}
\end{equation}
through the linear mapping
\begin{equation}
    s_{k} = \frac{s_{o} - s_{i}}{2} x_{k} + \frac{s_{o} + s_{i}}{2}.
\end{equation}
The resulting radial grid is non-uniform, with points clustering near both boundaries. The Fourier coefficients $f_{m}$ are then represented at each collocation point $s_k$ via the Chebyshev expansion
\begin{equation}
    f_{m}(s_{k}) \approx \left( \frac{2}{N_{s} - 1} \right)^{1/2}
    \sideset{}{''}\sum_{n = 0}^{N_{s} - 1} \hat{f}_{mn}\, T_{n}(x_{k}),
\end{equation}
where the double-prime summation indicates that the first and last terms are multiplied by $1/2$ \citep{glatzmaier1984numerical}, the hat denotes Chebyshev coefficients, and $T_{n}(x_{k})$ is the first-kind Chebyshev polynomial of order $n$ evaluated at $x_k$,
\begin{equation}
    T_{n}(x_k) \equiv \bm{T}_{kn} = \cos[n \arccos (x_{k})].
\end{equation}
Substituting the Fourier--Chebyshev expansions into \eqref{eq:omega-from-psi}, \eqref{eq:sph-qg-omega} and \eqref{eq:sph-qg-uphi-axisymmetric}, Eqs.~\eqref{eq:sph-qg-omega} and \eqref{eq:omega-from-psi} yield the following set of coupled semi-discrete equations for $\hat{\omega}_{mn}$ and $\hat{\psi}_{mn}$ for any $m > 0$,
\begin{align}
    \sideset{}{''}\sum_{n = 0}^{N_{s} - 1} \left[ \frac{\mathrm{d}}{\mathrm{d}t} \bm{T}_{kn} - \bm{O}_{kn}(m) \right] \hat{\omega}_{mn} &= \mathcal{N}_{\omega}(m, s_{k}), \label{appeq:omega-omega-system}\\
    \sideset{}{''}\sum_{n = 0}^{N_{s} - 1} \left[\bm{T}_{kn} \hat{\omega}_{mn} + \bm{P}_{kn}(m) \hat{\psi}_{mn} \right] &= 0. \label{appeq:omega-psi-system}
\end{align}
The real-valued collocation matrices entering \eqref{appeq:omega-omega-system} and \eqref{appeq:omega-psi-system} are
\begin{align}
    \bm{O}_{kn}(m) &= \bm{T}''_{kn} + \frac{1}{s_{k}} \bm{T}'_{kn} - \left( \frac{m^2}{s_{k}^2} + \Upsilon_{k} \right) \bm{T}_{kn}, \\
    \bm{P}_{kn}(m) &= \bm{T}''_{kn} + \left( \frac{1}{s_{k}} + \beta_{k} \right) \bm{T}'_{kn} + \left( \frac{\beta_{k}}{s_{k}} + \frac{\mathrm{d} \beta_{k}}{\mathrm{d} s} - \frac{m^2}{s_{k}^2} \right) \bm{T}_{kn},
\end{align}
where $\bm{T}''_{kn}$ and $\bm{T}'_{kn}$ are the second and first derivative matrices of the $n$-th order Chebyshev polynomial at $x_{k}$, respectively. The right-hand side in \eqref{appeq:omega-omega-system} comprises the prescribed forcing, the Coriolis force, and the nonlinear advection term, which is evaluated in physical space before being transformed back to spectral space,
\begin{equation}
    \mathcal{N}_{\omega}(m, s_{k}) = \mathcal{F}(m, s_{k}) - \frac{2}{E} \beta_{k} u_{s}(m, s_{k}) -\bm{\nabla \cdot} \left( \frac{1}{N_{\varphi}} \sum_{j = 0}^{N_{\varphi}} (\bm{u} \omega) \, e^{-\mathrm{ i } m \varphi_{j}} \right).
    \label{appeq:omega-non-linear}
\end{equation}
Likewise, Eq.~\eqref{eq:sph-qg-uphi-axisymmetric} for the mean azimuthal flow ${\langle u_\varphi \rangle_{\varphi}}$ is discretized as
\begin{equation}
    \sideset{}{''}\sum_{n = 0}^{N_{s} - 1} \left[ \frac{\mathrm{d}}{\mathrm{d}t} \bm{T}_{kn} - \bm{O}_{kn}(1) \right] \widehat{\langle u_\varphi \rangle_{\varphi}}_{n} = \mathcal{N}_{\langle u_\varphi \rangle_{\varphi}}(s_{k}),
\end{equation}
where the nonlinear right-hand side includes the self-interaction of the zonal wind \citep{aubert2003quasigeostrophic},
\begin{equation}
    \mathcal{N}_{\langle u_\varphi \rangle_{\varphi}}(s_{k}) = - \frac{E}{2} \Upsilon_{k} u_{\varphi_{0}} \omega_{0} - 2 \sum_{m = 1}^{N_{m}} \Re \left\{ u_{s_{m}} \omega_{m}^* \right\}.
    \label{appeq:uphi-axisymmetric-non-linear}
\end{equation}
Boundary conditions are imposed via the tau method: the first and last rows of each collocation matrix are replaced by the boundary constraints evaluated at $s = s_{o}$ and $s = s_{i}$, respectively. For a field $f$ subject to homogeneous Dirichlet conditions,
\begin{equation}
    \sideset{}{''}\sum_{n = 0}^{N_{s} - 1} \hat{f}_{mn} = 0 \mbox{ at } s = s_{o}, \qquad
    \sideset{}{''}\sum_{n = 0}^{N_{s} - 1} (-1)^{n} \hat{f}_{mn} = 0 \mbox{ at } s = s_{i}.
\end{equation}
For homogeneous Neumann conditions,
\begin{equation}
    \sideset{}{''}\sum_{n = 0}^{N_{s} - 1} n^{2} \hat{f}_{mn} = 0 \mbox{ at } s = s_{o}, \qquad
    \sideset{}{''}\sum_{n = 0}^{N_{s} - 1} (-1)^{n + 1} n^{2} \hat{f}_{mn} = 0 \mbox{ at } s = s_{i}.
\end{equation}
The coupled system \eqref{appeq:omega-omega-system} and \eqref{appeq:omega-psi-system} gives rise to the following $(2 N_{s} \times 2 N_{s})$ dense real matrix operator for each Fourier mode $m > 0$,
\begin{equation}
    \begin{bmatrix}
    \dfrac{\mathrm{d}}{\mathrm{d}t} \bm{T}_{kn} - \bm{O}_{kn} & 0 \\[6pt]
    \bm{T}_{kn} & \bm{P}_{kn}
    \end{bmatrix},
\end{equation}
with Dirichlet and Neumann boundary conditions on $\psi$ \eqref{eq:boundary-conditions}
imposed on the first and last rows of the top-right and bottom-right quadrants, respectively. The equation for the zonal flow $\langle u_\varphi \rangle_{\varphi}$ yields an independent $(N_{s} \times N_{s})$ operator with Dirichlet boundary conditions imposed on its first and last rows.

Collecting all spectral coefficients $(\hat{\omega}_{mn}, \hat{\psi}_{mn},
\widehat{\langle u_\varphi\rangle_\varphi}_n)$ into a single state vector
$\bm{z}$, the spatially discrete equations can be written in the compact additive split form
\begin{equation}
    \frac{\mathrm{d}\bm{z}}{\mathrm{d}t} =
    \mathcal{L}\bm{z} + \mathcal{N}(\bm{z}, t),
    \label{appeq:ode-split}
\end{equation}
where $\mathcal{L}$ denotes the stiff linear operator (viscous diffusion, Ekman pumping, and the streamfunction-vorticity coupling encoded in the collocation matrices $\bm{O}$ and $\bm{P}$), and $\mathcal{N}$ gathers all explicitly treated terms (the prescribed forcing $\mathcal{F}$, the Coriolis contribution, and the nonlinear advection terms in \eqref{appeq:omega-non-linear}--\eqref{appeq:uphi-axisymmetric-non-linear}). Any subgrid-scale correction $\mathcal{M}(\overline{\bm{z}}; \theta)$ is incorporated into $\mathcal{N}$ as an additive explicit source term.
Equation \eqref{appeq:ode-split} is advanced in time using an implicit-explicit Runge--Kutta (IMEX-RK) method, whose general structure is characterized by a double Butcher tableau $(\tilde{A}, \tilde{b}, c)$ / $(A, b, c)$ for the diagonally implicit (DIRK) and explicit (ERK) components, respectively \citep{boscarino2013implicit}. Starting from the state $\bm{z}^n$ at time $t^n$, the $s$-stage scheme first computes internal stage values $\bm{z}^{(i)}$, $i = 1, \ldots, s$, via
\begin{equation}
    \bm{z}^{(i)} = \bm{z}^{n} + \delta t \sum_{j=1}^{i-1} a_{ij} \, \mathcal{N} \! \left( \bm{z}^{(j)},\, t^{n} + c_j \delta t \right) + \delta t \sum_{j=1}^{i} \tilde{a}_{ij}\, \mathcal{L} \bm{z}^{(j)},
    \label{appeq:imex-stage}
\end{equation}
followed by the final update
\begin{equation}
    \bm{z}^{n+1} = \bm{z}^{n} + \delta t \sum_{i=1}^{s} b_i \, \mathcal{N} \! \left( \bm{z}^{(i)} \right) + \delta t \sum_{i=1}^{s} \tilde{b}_i\, \mathcal{L} \bm{z}^{(i)}.
    \label{appeq:imex-update}
\end{equation}
The ERK matrix $A = (a_{ij})$ is strictly lower triangular (so each $\mathcal{N}$ evaluation uses only previously computed stages), while the DIRK matrix $\tilde{A} = (\tilde{a}_{ij})$ is lower triangular with non-zero diagonal entries $\tilde{a}_{ii}$, which is the source of the implicit solve at each stage.
We use the BPR353 scheme of \citet[][Sec. 8.3]{boscarino2013implicit}, denoted BPR533 in the notation of \citet{gopinath2022assessment}, where the three integers represent the number of implicit DIRK stages ($s_I = 5$), the number of non-trivial explicit ERK stages ($s_E = 3$), and the order of accuracy ($r = 3$). The scheme is globally stiffly accurate, so that $\bm{z}^{n+1} \equiv \bm{z}^{(s)}$ and \eqref{appeq:imex-update} requires no separate assembly stage. Although the scheme involves $s_I = 5$ implicit stages, only $s_E = 3$ explicit evaluations of $\mathcal{N}$ are needed per time step; when the cost of the nonlinear transform dominates over that of the implicit solve ---as is typical in a pseudo-spectral setting--- this results in a competitive efficiency per time step relative to schemes with a larger number of explicit stages.
At each implicit stage, the linear system $(I - \delta t\,\tilde{a}_{ii} \, \mathcal{L}) \bm{z}^{(i)} = \cdots$ decouples in spectral space into independent per-mode systems involving the pre-factored $(2 N_s \times 2 N_s)$ and $(N_s \times N_s)$ collocation matrices; the LU factorizations are computed once at the start of the simulation and reused at every implicit stage of every time step. This scheme was selected over 25 competing IMEX integrators on the basis of the comprehensive assessment by \citet{gopinath2022assessment} in a similar pseudo-spectral context, where BPR353 was found to offer one of the best accuracy-to-cost ratios among third-order schemes.

\section{Power input \label{sec:power-input}}
In order to derive the expression of the input power $\power$, we have to uncurl the expression of the source of vorticity $\mathcal{F}$ in Eq.~\eqref{eq:forcing}. To that end, we begin by considering a single Gaussian pump located at the origin of the 2D plane. In cylindrical coordinates $(s,\varphi,z)$, we seek a vector $\mathbf{f}$ whose curl reads
\begin{equation}
    \bm{\nabla} \times \mathbf{f} = 
     \exp \left[ -\left( \frac{s}{\ell_{\mathcal F}} \right)^2 \right]     
    \bm{e}_z.
\end{equation}
Specifically, we require $\mathbf{f}$ to be of the form $\mathbf{f} = f(s) \bm{e}_\varphi$. In addition, we ask $f$ to be regular at $s = 0$. It is straightforward to obtain 
\begin{equation}
    f(s) = \frac{\ell_{\mathcal F}^2}{2 s} \left\{ 1 - \exp \left[ -\left( \frac{s}{\ell_{\mathcal F}} \right)^2\right] \right\}.
\end{equation}
To facilitate subsequent superposition, we switch from cylindrical to Cartesian coordinates, and note that 
\begin{equation}
    \bm{e}_\varphi = \frac{-y \bm{e}_x + x \bm{e}_y}{s}. 
\end{equation}
Consequently, for a single unitary Gaussian source located at the origin, we have
\begin{equation}
    \mathbf{f} = f(s) \bm{e}_\varphi = \frac{\ell_{\mathcal F}^2}{2 s^2} \left\{ 1 - \left[\exp - \left( \frac{s}{\ell_{\mathcal F}} \right)^2 \right]
    \right\}\left( -y \bm{e}_x + x \bm{e}_y \right).
\end{equation}
If the center of the source is now placed at $(x_i,y_i)$ the previous expression becomes 
\begin{equation}
    \mathbf{f} = \frac{\ell_{\mathcal F}^2}{2 s_i^2} \left\{ 1 - \exp \left[ -\left( \frac{s_i}{\ell_{\mathcal F}} \right)^2 \right] \right\} \left[ -(y - y_i) \bm{e}_x + (x - x_i) \bm{e}_y \right],
\label{eq:singlepump}
\end{equation}
where $s_i = s_i(x,y)=\sqrt{\left( x - x_i \right)^2 + \left( y - y_i \right)^2}$. 
Let $p_{\mathcal{F}}$ denote the input power density. Using Eq.~\eqref{eq:singlepump} in conjunction with Eq.~\eqref{eq:forcing} leads by superposition to  
\begin{equation}
    p_\mathcal{F} = \dfrac{a_{\mathcal{F}} \ell_{\mathcal{F}}^2}{2} \sum_{i = 1}^{N} \frac{(-1)^{i}}{s_i^2} \left\{ 1 - \exp \left[ - \left( \frac{s_{i}}{\ell_{\mathcal{F}}} \right)^{2} \right] \right\}\left[ (x - x_i) u_y - (y - y_i) u_x) \right],
\end{equation}
where $u_x = u_s \cos \varphi - u_\varphi \sin \varphi$ and $u_y = u_s\sin \varphi + u_\varphi \cos\varphi$. The input power at any time $t$ averaged over the domain is then $\power = \langle p_\mathcal{F} \rangle$, with the domain average operator defined in Eq.~\eqref{eq:def_average}.

\bibliography{apssamp}

@article{landeau2022sustaining,
  title={Sustaining {E}arth’s magnetic dynamo},
  author={Landeau, Maylis and Fournier, Alexandre and Nataf, Henri-Claude and C{\'e}bron, David and Schaeffer, Nathana{\"e}l},
  journal={Nature Reviews Earth \& Environment},
  volume={3},
  number={4},
  pages={255--269},
  year={2022},
  publisher={Nature Publishing Group UK London},
  doi={10.1038/s43017-022-00264-1}
}

@article{finlay2023gyres,
  title={Gyres, jets and waves in the {E}arth’s core},
  author={Finlay, Christopher C and Gillet, Nicolas and Aubert, Julien and Livermore, Philip W and Jault, Dominique},
  journal={Nature Reviews Earth \& Environment},
  volume={4},
  number={6},
  pages={377--392},
  year={2023},
  publisher={Nature Publishing Group UK London},
  doi={10.1038/s43017-023-00425-w}
}

@article{pais2008quasi,
  title={Quasi-geostrophic flows responsible for the secular variation of the {E}arth's magnetic field},
  author={Pais, MA and Jault, Dominique},
  journal={Geophysical Journal International},
  volume={173},
  number={2},
  pages={421--443},
  year={2008},
  publisher={Blackwell Publishing Ltd Oxford, UK},
  doi={10.1111/j.1365-246X.2008.03741.x}
}

@article{glatzmaier1995three,
  title={A three-dimensional self-consistent computer simulation of a geomagnetic field reversal},
  author={Glatzmaier, Gary A and Roberts, Paul H},
  journal={Nature},
  volume={377},
  number={6546},
  pages={203--209},
  year={1995},
  publisher={Nature Publishing Group UK London},
  doi={10.1038/377203a0}
}

@article{wicht2016gaussian,
  title={A Gaussian model for simulated geomagnetic field reversals},
  author={Wicht, Johannes and Meduri, Domenico G},
  journal={Physics of the Earth and Planetary Interiors},
  volume={259},
  pages={45--60},
  year={2016},
  publisher={Elsevier},
  doi={10.1016/j.pepi.2016.07.007}
}

@article{smagorinsky1963general,
  title={General circulation experiments with the primitive equations: I. The basic experiment},
  author={Smagorinsky, Joseph},
  journal={Monthly Weather Review},
  volume={91},
  number={3},
  pages={99--164},
  year={1963},
  doi={10.1175/1520-0493(1963)091<0099:GCEWTP>2.3.CO;2}
}

@article{maltrud1991energy,
  title={Energy spectra and coherent structures in forced two-dimensional and beta-plane turbulence},
  author={Maltrud, ME and Vallis, GK},
  journal={Journal of Fluid Mechanics},
  volume={228},
  pages={321--342},
  year={1991},
  publisher={Cambridge University Press},
  doi={10.1017/S0022112091002720}
}

@article{chen2005large,
  title={Large eddy simulations of two-dimensional turbulent convection in a density-stratified fluid},
  author={Chen, Qiaoning and Glatzmaier, Gary A},
  journal={Geophysical and Astrophysical Fluid Dynamics},
  volume={99},
  number={5},
  pages={355--375},
  year={2005},
  publisher={Taylor \& Francis},
  doi={10.1080/03091920500264513}
}

@article{kuang1999numerical,
  title={Numerical modeling of magnetohydrodynamic convection in a rapidly rotating spherical shell: weak and strong field dynamo action},
  author={Kuang, Weijia and Bloxham, Jeremy},
  journal={Journal of Computational Physics},
  volume={153},
  number={1},
  pages={51--81},
  year={1999},
  publisher={Elsevier},
  doi={10.1006/jcph.1999.6274}
}

@book{nataf2015turbulence,
  title={Turbulence in the core},
  author={Nataf, H-C and Schaeffer, Nathana{\"e}l},
  booktitle={Treatise on Geophysics},
  edition={2},
  volume={8},
  pages={161--181},
  year={2015},
  publisher={Elsevier},
  address={Amsterdam},
  editor={Schubert, G},
  doi={10.1016/B978-0-444-53802-4.00142-1}
}

@article{aubert2017spherical,
  title={Spherical convective dynamos in the rapidly rotating asymptotic regime},
  author={Aubert, Julien and Gastine, Thomas and Fournier, Alexandre},
  journal={Journal of Fluid Mechanics},
  volume={813},
  pages={558--593},
  year={2017},
  publisher={Cambridge University Press},
  doi={10.1017/jfm.2016.789}
}

@article{germano1986proposal,
  title={A proposal for a redefinition of the turbulent stresses in the filtered {N}avier-{S}tokes equations},
  author={Germano, Massimo},
  journal={Physics of Fluids},
  volume={29},
  number={7},
  pages={2323--2324},
  year={1986},
  publisher={AIP Publishing},
  doi={10.1063/1.865568}
}

@incollection{leonard1975energy,
  title={Energy cascade in large-eddy simulations of turbulent fluid flows},
  author={Leonard, Anthony},
  booktitle={Advances in Geophysics},
  volume={18},
  pages={237--248},
  year={1975},
  publisher={Academic Press},
  address={New York, NY, USA},
  editor={Frenkiel, F N and Munn, R E},
  doi={10.1016/S0065-2687(08)60464-1}
}

@inproceedings{bardina1980improved,
  title={Improved subgrid-scale models for large-eddy simulation},
  author={Bardina, Jorge and Ferziger, J and Reynolds, W C},
  booktitle={13th Fluid and Plasma Dynamics Conference},
  address={Snowmass, CO, USA},
  pages={1357},
  year={1980},
  publisher={American Institute of Aeronautics and Astronautics},
  doi={10.2514/6.1980-1357}
}

@article{buffett2003comparison,
  title={A comparison of subgrid-scale models for large-eddy simulations of convection in the {E}arth's core},
  author={Buffett, Bruce A},
  journal={Geophysical Journal International},
  volume={153},
  number={3},
  pages={753--765},
  year={2003},
  publisher={Blackwell Science Ltd Oxford, UK},
  doi={10.1046/j.1365-246X.2003.01930.x}
}

@article{matsui2005sub,
  title={Sub-grid scale model for convection-driven dynamos in a rotating plane layer},
  author={Matsui, Hiroaki and Buffett, Bruce A},
  journal={Physics of the Earth and Planetary Interiors},
  volume={153},
  number={1-3},
  pages={108--123},
  year={2005},
  publisher={Elsevier},
  doi={10.1016/j.pepi.2005.03.019}
}

@article{matsui2007commutation,
  title={Commutation error correction for large eddy simulations of convection driven dynamos},
  author={Matsui, Hiroaki and Buffett, Bruce A},
  journal={Geophysical and Astrophysical Fluid Dynamics},
  volume={101},
  number={5-6},
  pages={429--449},
  year={2007},
  publisher={Taylor \& Francis},
  doi={10.1080/03091920701562129}
}

@article{matsui2007dynamic,
  title={A dynamic scale-similarity model for dynamo simulations in a rotating plane layer},
  author={Matsui, Hiroaki and Buffett, Bruce A},
  journal={Geophysical and Astrophysical Fluid Dynamics},
  volume={101},
  number={5-6},
  pages={451--468},
  year={2007},
  publisher={Taylor \& Francis},
  doi={10.1080/03091920701562145}
}

@article{matsui2012large,
  title={Large-eddy simulations of convection-driven dynamos using a dynamic scale-similarity model},
  author={Matsui, Hiroaki and Buffett, Bruce A},
  journal={Geophysical \& Astrophysical Fluid Dynamics},
  volume={106},
  number={3},
  pages={250--276},
  year={2012},
  publisher={Taylor \& Francis},
  doi={10.1080/03091929.2011.590806}
}

@article{brunton2020machine,
  title={Machine learning for fluid mechanics},
  author={Brunton, Steven L and Noack, Bernd R and Koumoutsakos, Petros},
  journal={Annual Review of Fluid Mechanics},
  volume={52},
  number={1},
  pages={477--508},
  year={2020},
  publisher={Annual Reviews},
  doi={10.1146/annurev-fluid-010719-060214}
}

@article{vinuesa2022enhancing,
  title={Enhancing computational fluid dynamics with machine learning},
  author={Vinuesa, Ricardo and Brunton, Steven L},
  journal={Nature Computational Science},
  volume={2},
  number={6},
  pages={358--366},
  year={2022},
  publisher={Nature Publishing Group US New York},
  doi={10.1038/s43588-022-00264-7}
}

@article{yuval2020stable,
  title={Stable machine-learning parameterization of subgrid processes for climate modeling at a range of resolutions},
  author={Yuval, Janni and O’Gorman, Paul A},
  journal={Nature communications},
  volume={11},
  number={1},
  pages={3295},
  year={2020},
  publisher={Nature Publishing Group UK London},
  doi={10.1038/s41467-020-17142-3}
}

@article{arcomano2023hybrid,
  title={A hybrid atmospheric model incorporating machine learning can capture dynamical processes not captured by its physics-based component},
  author={Arcomano, Troy and Szunyogh, Istvan and Wikner, Alexander and Hunt, Brian R and Ott, Edward},
  journal={Geophysical Research Letters},
  volume={50},
  number={8},
  pages={e2022GL102649},
  year={2023},
  publisher={Wiley Online Library},
  doi={10.1029/2022GL102649}
}

@article{bolton2019applications,
  title={Applications of deep learning to ocean data inference and subgrid parameterization},
  author={Bolton, Thomas and Zanna, Laure},
  journal={Journal of Advances in Modeling Earth Systems},
  volume={11},
  number={1},
  pages={376--399},
  year={2019},
  publisher={Wiley Online Library},
  doi={10.1029/2018MS001472}
}

@article{finn2023deep,
  title={Deep learning subgrid-scale parametrisations for short-term forecasting of sea-ice dynamics with a {M}axwell elasto-brittle rheology},
  author={Finn, Tobias Sebastian and Durand, Charlotte and Farchi, Alban and Bocquet, Marc and Chen, Yumeng and Carrassi, Alberto and Dansereau, V{\'e}ronique},
  journal={The Cryosphere},
  volume={17},
  number={7},
  pages={2965--2991},
  year={2023},
  publisher={Copernicus Publications G{\"o}ttingen, Germany},
  doi={10.5194/tc-17-2965-2023}
}

@article{rasp2018deep,
  title={Deep learning to represent subgrid processes in climate models},
  author={Rasp, Stephan and Pritchard, Michael S and Gentine, Pierre},
  journal={Proceedings of the national academy of sciences},
  volume={115},
  number={39},
  pages={9684--9689},
  year={2018},
  publisher={National Academy of Sciences},
  doi={10.1073/pnas.1810286115}
}

@article{dueben2018challenges,
  title={Challenges and design choices for global weather and climate models based on machine learning},
  author={Dueben, Peter D and Bauer, Peter},
  journal={Geoscientific Model Development},
  volume={11},
  number={10},
  pages={3999--4009},
  year={2018},
  publisher={Copernicus GmbH},
  doi={10.5194/gmd-11-3999-2018}
}

@article{yuval2021use,
  title={Use of neural networks for stable, accurate and physically consistent parameterization of subgrid atmospheric processes with good performance at reduced precision},
  author={Yuval, Janni and O'Gorman, Paul A and Hill, Chris N},
  journal={Geophysical Research Letters},
  volume={48},
  number={6},
  pages={e2020GL091363},
  year={2021},
  publisher={Wiley Online Library},
  doi={10.1029/2020GL091363}
}

@article{gelbrecht2023differentiable,
  title={Differentiable programming for {E}arth system modeling},
  author={Gelbrecht, Maximilian and White, Alistair and Bathiany, Sebastian and Boers, Niklas},
  journal={Geoscientific Model Development},
  volume={16},
  number={11},
  pages={3123--3135},
  year={2023},
  publisher={Copernicus Publications G{\"o}ttingen, Germany},
  doi={10.5194/gmd-16-3123-2023}
}

@article{rasp2020coupled,
  title={Coupled online learning as a way to tackle instabilities and biases in neural network parameterizations: General algorithms and {L}orenz 96 case study (v1. 0)},
  author={Rasp, Stephan},
  journal={Geoscientific Model Development},
  volume={13},
  number={5},
  pages={2185--2196},
  year={2020},
  publisher={Copernicus Publications G{\"o}ttingen, Germany},
  doi={10.5194/gmd-13-2185-2020}
}

@article{kochkov2021machine,
  title={Machine learning--accelerated computational fluid dynamics},
  author={Kochkov, Dmitrii and Smith, Jamie A and Alieva, Ayya and Wang, Qing and Brenner, Michael P and Hoyer, Stephan},
  journal={Proceedings of the National Academy of Sciences},
  volume={118},
  number={21},
  pages={e2101784118},
  year={2021},
  publisher={National Academy of Sciences},
  doi={10.1073/pnas.2101784118}
}

@article{frezat2022posteriori,
  title={A posteriori learning for quasi-geostrophic turbulence parametrization},
  author={Frezat, Hugo and Le Sommer, Julien and Fablet, Ronan and Balarac, Guillaume and Lguensat, Redouane},
  journal={Journal of Advances in Modeling Earth Systems},
  volume={14},
  number={11},
  pages={e2022MS003124},
  year={2022},
  publisher={Wiley Online Library},
  doi={10.1029/2022MS003124}
}

@article{yan2025adjoint,
  title={Adjoint-based online learning of two-layer quasi-geostrophic baroclinic turbulence},
  author={Yan, Fei Er and Frezat, Hugo and Sommer, Julien Le and Mak, Julian and Otness, Karl},
  journal={Journal of Advances in Modeling Earth Systems},
  volume={17},
  number={7},
  pages={e2024MS004857},
  year={2025},
  publisher={Wiley Online Library},
  doi={10.1029/2024MS004857}
}

@article{kochkov2024neural,
  title={Neural general circulation models for weather and climate},
  author={Kochkov, Dmitrii and Yuval, Janni and Langmore, Ian and Norgaard, Peter and Smith, Jamie and Mooers, Griffin and Kl{\"o}wer, Milan and Lottes, James and Rasp, Stephan and D{\"u}ben, Peter and others},
  journal={Nature},
  volume={632},
  number={8027},
  pages={1060--1066},
  year={2024},
  publisher={Nature Publishing Group UK London},
  doi={10.1038/s41586-024-07744-y}
}

@article{lemasquerier2023zonal,
  title={Zonal jets experiments in the gas giants’ zonostrophic regime},
  author={Lemasquerier, Daphn{\'e} and Favier, Benjamin and Le Bars, Michael},
  journal={Icarus},
  volume={390},
  pages={115292},
  year={2023},
  publisher={Elsevier},
  doi={10.1016/j.icarus.2022.115292}
}

@article{rotvig2007multiple,
  title={Multiple zonal jets and drifting: thermal convection in a rapidly rotating spherical shell compared to a quasigeostrophic model},
  author={Rotvig, Jon},
  journal={Physical Review E—Statistical, Nonlinear, and Soft Matter Physics},
  volume={76},
  number={4},
  pages={046306},
  year={2007},
  publisher={APS},
  doi={10.1103/PhysRevE.76.046306}
}

@article{chemke2015poleward,
  title={Poleward migration of eddy-driven jets},
  author={Chemke, Rei and Kaspi, Yohai},
  journal={Journal of Advances in Modeling Earth Systems},
  volume={7},
  number={3},
  pages={1457--1471},
  year={2015},
  publisher={Wiley Online Library},
  doi={10.1002/2015MS000481}
}

@article{lonner2022planetary,
  title={Planetary core-style rotating convective flows in paraboloidal laboratory experiments},
  author={Lonner, Taylor L and Aggarwal, Ashna and Aurnou, Jonathan M},
  journal={Journal of Geophysical Research: Planets},
  volume={127},
  number={10},
  pages={e2022JE007356},
  year={2022},
  publisher={Wiley Online Library},
  doi={10.1029/2022JE007356}
}

@phdthesis{cope2021dynamics,
  title={The dynamics of geophysical and astrophysical turbulence},
  author={Cope, Laura},
  school={University of Cambridge},
  year={2021},
  doi={10.17863/CAM.75705}
}

@article{cabanes2017laboratory,
  title={A laboratory model for deep-seated jets on the gas giants},
  author={Cabanes, Simon and Aurnou, Jonathan and Favier, Benjamin and Le Bars, Michael},
  journal={Nature Physics},
  volume={13},
  number={4},
  pages={387--390},
  year={2017},
  publisher={Nature Publishing Group UK London},
  doi={10.1038/nphys4001}
}

@article{zhu2025zonal,
  title={Zonal jets over small-scale topography: jet properties and energy transfers},
  author={Zhu, Hang-Yu and Pei, Tian-Yi and Xie, Jin-Han and Xia, Ke-Qing},
  journal={Journal of Fluid Mechanics},
  volume={1020},
  pages={A35},
  year={2025},
  publisher={Cambridge University Press},
  doi={10.1017/jfm.2025.10566 }
}

@article{aubert2003quasigeostrophic,
  title={Quasigeostrophic models of convection in rotating spherical shells},
  author={Aubert, Julien and Gillet, Nicolas and Cardin, Philippe},
  journal={Geochemistry, Geophysics, Geosystems},
  volume={4},
  number={7},
  year={2003},
  publisher={Wiley Online Library},
  doi={10.1029/2002GC000456}
}

@article{guervilly2017multiple,
  title={Multiple zonal jets and convective heat transport barriers in a quasi-geostrophic model of planetary cores},
  author={Guervilly, C{\'e}line and Cardin, Philippe},
  journal={Geophysical Journal International},
  volume={211},
  number={1},
  pages={455--471},
  year={2017},
  publisher={Oxford University Press},
  doi={10.1093/gji/ggx315}
}

@article{gillet2006quasi,
  title={The quasi-geostrophic model for rapidly rotating spherical convection outside the tangent cylinder},
  author={Gillet, N and Jones, CA},
  journal={Journal of Fluid Mechanics},
  volume={554},
  pages={343--369},
  year={2006},
  publisher={Cambridge University Press},
  doi={10.1017/S0022112006009219}
}

@article{gastine2019pizza,
  title={pizza: an open-source pseudo-spectral code for spherical quasi-geostrophic convection},
  author={Gastine, Thomas},
  journal={Geophysical Journal International},
  volume={217},
  number={3},
  pages={1558--1576},
  year={2019},
  publisher={Oxford University Press},
  doi={10.1093/gji/ggz103}
}

@article{plaut2002low,
  title={Low-{P}randtl-number convection in a rotating cylindrical annulus},
  author={Plaut, Emmanuel and Busse, Friedrich H},
  journal={Journal of Fluid Mechanics},
  volume={464},
  pages={345--363},
  year={2002},
  publisher={Cambridge University Press},
  doi={10.1017/S0022112002008923}
}

@book{greenspan1968theory,
  title={The {T}heory of {R}otating {F}luids},
  author={Greenspan, Harvey Philip},
  year={1968},
  publisher={Cambridge University Press},
  address={Cambridge}
}

@article{schaeffer2005quasigeostrophic,
  title={Quasigeostrophic model of the instabilities of the {S}tewartson layer in flat and depth-varying containers},
  author={Schaeffer, Nathana{\"e}l and Cardin, Philippe},
  journal={Physics of Fluids},
  volume={17},
  number={10},
  year={2005},
  publisher={AIP Publishing},
  doi={10.1063/1.2073547}
}

@article{boscarino2013implicit,
  title={Implicit-explicit {R}unge--{K}utta schemes for hyperbolic systems and kinetic equations in the diffusion limit},
  author={Boscarino, Sebastiano and Pareschi, Lorenzo and Russo, Giovanni},
  journal={{SIAM} Journal on Scientific Computing},
  volume={35},
  number={1},
  pages={A22--A51},
  year={2013},
  publisher={SIAM},
  doi={10.1137/110842855}
}

@article{gopinath2022assessment,
  title={An assessment of implicit-explicit time integrators for the pseudo-spectral approximation of {B}oussinesq thermal convection in an annulus},
  author={Gopinath, Venkatesh and Fournier, Alexandre and Gastine, Thomas},
  journal={Journal of Computational Physics},
  volume={460},
  pages={110965},
  year={2022},
  publisher={Elsevier},
  doi={10.1016/j.jcp.2022.110965}
}

@article{maulik2019subgrid,
  title={Subgrid modelling for two-dimensional turbulence using neural networks},
  author={Maulik, Romit and San, Omer and Rasheed, Adil and Vedula, Prakash},
  journal={Journal of Fluid Mechanics},
  volume={858},
  pages={122--144},
  year={2019},
  publisher={Cambridge University Press},
  doi={10.1017/jfm.2018.770 }
}

@article{guan2022stable,
  title={Stable a posteriori {LES} of 2{D} turbulence using convolutional neural networks: Backscattering analysis and generalization to higher {R}e via transfer learning},
  author={Guan, Yifei and Chattopadhyay, Ashesh and Subel, Adam and Hassanzadeh, Pedram},
  journal={Journal of Computational Physics},
  volume={458},
  pages={111090},
  year={2022},
  publisher={Elsevier},
  doi={10.1016/j.jcp.2022.111090}
}

@article{sirignano2020dpm,
  title={{DPM}: A deep learning {PDE} augmentation method with application to large-eddy simulation},
  author={Sirignano, Justin and MacArt, Jonathan F and Freund, Jonathan B},
  journal={Journal of Computational Physics},
  volume={423},
  pages={109811},
  year={2020},
  publisher={Elsevier},
  doi={10.1016/j.jcp.2020.109811}
}

@article{frezat2021physical,
  title={Physical invariance in neural networks for subgrid-scale scalar flux modeling},
  author={Frezat, Hugo and Balarac, Guillaume and Le Sommer, Julien and Fablet, Ronan and Lguensat, Redouane},
  journal={Physical Review Fluids},
  volume={6},
  number={2},
  pages={024607},
  year={2021},
  publisher={APS},
  doi={10.1103/PhysRevFluids.6.024607}
}

@article{zhou2019subgrid,
  title={Subgrid-scale model for large-eddy simulation of isotropic turbulent flows using an artificial neural network},
  author={Zhou, Zhideng and He, Guowei and Wang, Shizhao and Jin, Guodong},
  journal={Computers \& Fluids},
  volume={195},
  pages={104319},
  year={2019},
  publisher={Elsevier},
  doi={10.1016/j.compfluid.2019.104319}
}

@article{shen1995efficient,
  title={Efficient spectral-{G}alerkin method {II}. Direct solvers of second-and fourth-order equations using {C}hebyshev polynomials},
  author={Shen, Jie},
  journal={SIAM Journal on Scientific Computing},
  volume={16},
  number={1},
  pages={74--87},
  year={1995},
  publisher={SIAM},
  doi={10.1137/0915089}
}

@article{julien2009efficient,
  title={Efficient multi-dimensional solution of {PDE}s using {C}hebyshev spectral methods},
  author={Julien, Keith and Watson, Mike},
  journal={Journal of Computational Physics},
  volume={228},
  number={5},
  pages={1480--1503},
  year={2009},
  publisher={Elsevier},
  doi={10.1016/j.jcp.2008.10.043}
}

@article{sanderse2025scientific,
  title={Scientific machine learning for closure models in multiscale problems: A review},
  author={Sanderse, Benjamin and Stinis, Panos and Maulik, Romit and Ahmed, Shady E},
  journal={Foundations of Data Science},
  volume={7},
  number={1},
  pages={298-337},
  year={2025},
  doi={10.3934/fods.2024043}
}

@book{sagaut2006large,
  title={Large {E}ddy {S}imulation for {I}ncompressible {F}lows},
  author={Sagaut, Pierre},
  year={2006},
  publisher={Springer},
  address={Berlin, Heidelberg},
  doi={10.1007/b137536}
}

@article{ghosal1995basic,
  title={The basic equations for the large eddy simulation of turbulent flows in complex geometry},
  author={Ghosal, Sandip and Moin, Parviz},
  journal={Journal of Computational Physics},
  volume={118},
  number={1},
  pages={24--37},
  year={1995},
  publisher={Elsevier},
  doi={10.1006/jcph.1995.1077}
}

@article{yalla2021effects,
  title={Effects of resolution inhomogeneity in large-eddy simulation},
  author={Yalla, Gopal R and Oliver, Todd A and Haering, Sigfried W and Engquist, Bj{\"o}rn and Moser, Robert D},
  journal={Physical Review Fluids},
  volume={6},
  number={7},
  pages={074604},
  year={2021},
  publisher={APS},
  doi={10.1103/PhysRevFluids.6.074604}
}

@article{jakhar2024learning,
  title={Learning closed-form equations for subgrid-scale closures from high-fidelity data: Promises and challenges},
  author={Jakhar, Karan and Guan, Yifei and Mojgani, Rambod and Chattopadhyay, Ashesh and Hassanzadeh, Pedram},
  journal={Journal of Advances in Modeling Earth Systems},
  volume={16},
  number={7},
  pages={e2023MS003874},
  year={2024},
  publisher={Wiley Online Library},
  doi={10.1029/2023MS003874}
}

@article{ouala2024online,
  title={Online calibration of deep learning sub-models for hybrid numerical modeling systems},
  author={Ouala, Said and Chapron, Bertrand and Collard, Fabrice and Gaultier, Lucile and Fablet, Ronan},
  journal={Communications Physics},
  volume={7},
  number={1},
  pages={402},
  year={2024},
  publisher={Nature Publishing Group UK London},
  doi={10.1038/s42005-024-01880-7}
}

@article{frezat2026gradient,
  title={Gradient-free online learning of subgrid-scale dynamics with neural emulators},
  author={Frezat, Hugo and Fablet, Ronan and Balarac, Guillaume and Le Sommer, Julien},
  journal={Journal of Advances in Modeling Earth Systems},
  volume={18},
  number={8},
  pages={e2023MS004140},
  year={2026},
  publisher={Wiley Online Library},
  doi={10.1029/2023MS004140}
}

@misc{jax2018github,
  title={{JAX}: composable transformations of {P}ython+{N}um{P}y programs},
  author={James Bradbury and Roy Frostig and Peter Hawkins and Matthew James Johnson and Chris Leary and Dougal Maclaurin and George Necula and Adam Paszke and Jake Vander{P}las and Skye Wanderman-{M}ilne and Qiao Zhang},
  version={0.3.13},
  year={2018},
  note={http://github.com/jax-ml/jax}
}

@article{baydin2018automatic,
  title={Automatic differentiation in machine learning: a survey},
  author={Baydin, Atilim Gunes and Pearlmutter, Barak A and Radul, Alexey Andreyevich and Siskind, Jeffrey Mark},
  journal={J. Mach. Learn. Res.},
  volume={18},
  number={1},
  pages={5595–-5637},
  year={2017}
}

@article{sapienza2024differentiable,
  title={Differentiable Programming for Differential Equations: A Review},
  author={Sapienza, Facundo and Bolibar, Jordi and Sch{\"a}fer, Frank and Groenke, Brian and Pal, Avik and Boussange, Victor and Heimbach, Patrick and Hooker, Giles and P{\'e}rez, Fernando and Persson, Per-Olof and others},
  journal={arXiv preprint},
  year={2024},
  doi={10.48550/arXiv.2406.09699}
}

@inproceedings{loshchilov2018decoupled,
  title={Decoupled Weight Decay Regularization},
  author={Ilya Loshchilov and Frank Hutter},
  booktitle={7th International Conference on Learning Representations (ICLR 2019)},
  address={New Orleans, LA, USA},
  publisher={Curran Associates, Inc.},
  year={2019},
  url={https://openreview.net/forum?id=Bkg6RiCqY7},
}

@book{hirose2006complex,
  title={Complex-{V}alued {N}eural {N}etworks},
  author={Hirose, Akira},
  year={2006},
  publisher={Springer},
  address={Berlin, Heidelberg},
  doi={10.1007/978-3-540-33457-6}
}

@inproceedings{trabelsi2018deep,
  title={Deep Complex Networks},
  author={Chiheb Trabelsi and Olexa Bilaniuk and Ying Zhang and Dmitriy Serdyuk and Sandeep Subramanian and Joao Felipe Santos and Soroush Mehri and Negar Rostamzadeh and Yoshua Bengio and Christopher J Pal},
  booktitle={6th International Conference on Learning Representations (ICLR 2018)},
  address={Vancouver, BC, Canada},
  publisher={Curran Associates, Inc.},
  year={2018},
  url={https://openreview.net/forum?id=H1T2hmZAb},
}

@inproceedings{liu2022convnet,
  title={A {C}onv{N}et for the 2020s},
  author={Liu, Zhuang and Mao, Hanzi and Wu, Chao-Yuan and Feichtenhofer, Christoph and Darrell, Trevor and Xie, Saining},
  booktitle={Proceedings of the IEEE/CVF Conference on Computer Vision and Pattern Recognition (CVPR)},
  address={New Orleans, LA, USA},
  pages={11976--11986},
  publisher={IEEE},
  year={2022},
  url={https://openreview.net/forum?id=yXxBaGaQEh}
}

@inproceedings{arjovsky2016unitary,
  title={Unitary evolution recurrent neural networks},
  author={Arjovsky, Martin and Shah, Amar and Bengio, Yoshua},
  booktitle={Proceedings of the 33rd International Conference on Machine Learning},
  address={New York, NY, USA},
  pages={1120--1128},
  publisher={PMLR},
  year={2016},
  url={https://openreview.net/forum?id=71BmDZPEluAE8VvKUQ80}
}

@article{hjorungnes2007complex,
  title={Complex-valued matrix differentiation: Techniques and key results},
  author={Hj{\o}rungnes, Are and Gesbert, David},
  journal={IEEE Transactions on Signal Processing},
  volume={55},
  number={6},
  pages={2740--2746},
  year={2007},
  publisher={IEEE},
  doi={10.1109/TSP.2007.893762}
}

@article{shishkina2010boundary,
  title={Boundary layer structure in turbulent thermal convection and its consequences for the required numerical resolution},
  author={Shishkina, Olga and Stevens, Richard JAM and Grossmann, Siegfried and Lohse, Detlef},
  journal={New journal of Physics},
  volume={12},
  number={7},
  pages={075022},
  year={2010},
  publisher={IOP Publishing},
  doi={10.1088/1367-2630/12/7/075022}
}

@article{barrois2022comparison,
  title={Comparison of quasi-geostrophic, hybrid and 3-{D} models of planetary core convection},
  author={Barrois, Olivier and Gastine, Thomas and Finlay, Christopher C},
  journal={Geophysical Journal International},
  volume={231},
  number={1},
  pages={129--158},
  year={2022},
  publisher={Oxford University Press},
  doi={10.1093/gji/ggac141}
}

@article{davy2026effect,
  title={Effect of hyperdiffusion on rotating {R}ayleigh-{B}{\'e}nard convection},
  author={Davy, B and Davies, C J and Mound, J E and Tobias, S M},
  journal={Physical Review Fluids},
  volume={11},
  number={1},
  pages={013502},
  year={2026},
  publisher={APS},
  doi={10.1103/jvgr-d2mw}
}

@article{leith1996stochastic,
  title={Stochastic models of chaotic systems},
  author={Leith, CE},
  journal={Physica D: Nonlinear Phenomena},
  volume={98},
  number={2-4},
  pages={481--491},
  year={1996},
  publisher={Elsevier},
  doi={10.1016/0167-2789(96)00107-8}
}

@article{fox2008can,
  title={Can large eddy simulation techniques improve mesoscale rich ocean models?},
  author={Fox-Kemper, Baylor and Menemenlis, Dimitris},
  journal={Geophysical Monograph Series},
  volume={177},
  pages={319--337},
  year={2008},
  doi={10.1029/177GM19}
}

@article{bachman2017scale,
  title={A scale-aware subgrid model for quasi-geostrophic turbulence},
  author={Bachman, Scott D and Fox-Kemper, Baylor and Pearson, Brodie},
  journal={Journal of Geophysical Research: Oceans},
  volume={122},
  number={2},
  pages={1529--1554},
  year={2017},
  publisher={Wiley Online Library},
  doi={10.1002/2016JC012265}
}

@article{grooms2023backscatter,
  title={Backscatter in energetically-constrained {L}eith parameterizations},
  author={Grooms, Ian},
  journal={Ocean Modelling},
  volume={186},
  pages={102265},
  year={2023},
  publisher={Elsevier},
  doi={10.1016/j.ocemod.2023.102265}
}

@article{wilder2025examining,
  title={Examining the fidelity of {L}eith subgrid closures for parameterizing mesoscale eddies in idealized and global ({NEMO}) ocean models},
  author={Wilder, T and Kuhlbrodt, T},
  journal={Journal of Advances in Modeling Earth Systems},
  volume={17},
  number={9},
  pages={e2025MS004950},
  year={2025},
  publisher={Wiley Online Library},
  doi={10.1029/2025MS004950}
}

@article{rhines1975waves,
  title={Waves and turbulence on a beta-plane},
  author={Rhines, Peter B},
  journal={Journal of Fluid Mechanics},
  volume={69},
  number={3},
  pages={417--443},
  year={1975},
  publisher={Cambridge University Press},
  doi={10.1017/S0022112075001504}
}

@article{heimpel2007turbulent,
  title={Turbulent convection in rapidly rotating spherical shells: A model for equatorial and high latitude jets on {J}upiter and {S}aturn},
  author={Heimpel, Moritz and Aurnou, Jonathan},
  journal={Icarus},
  volume={187},
  number={2},
  pages={540--557},
  year={2007},
  publisher={Elsevier},
  doi={10.1016/j.icarus.2006.10.023}
}

@incollection{mcintyre2003potential,
  title={Potential vorticity},
  author={McIntyre, Michael E},
  booktitle={Encyclopedia of Atmospheric Sciences},
  volume={2},
  pages={685--694},
  year={2003},
  publisher={Academic Press},
  address={Oxford},
  editor={Holton, James R and Pyle, John A and Curry, Judith A},
  doi={10.1016/B0-12-227090-8/00484-X}
}

@article{scott2012structure,
  title={The structure of zonal jets in geostrophic turbulence},
  author={Scott, Richard K and Dritschel, David G},
  journal={Journal of Fluid Mechanics},
  volume={711},
  pages={576--598},
  year={2012},
  publisher={Cambridge University Press},
  doi={10.1017/jfm.2012.410}
}

@article{verhoeven2014compressional,
  title={The compressional beta effect: A source of zonal winds in planets?},
  author={Verhoeven, Jan and Stellmach, Stephan},
  journal={Icarus},
  volume={237},
  pages={143--158},
  year={2014},
  publisher={Elsevier},
  doi={10.1016/j.icarus.2014.04.019}
}

@article{sukoriansky2002universal,
  title={Universal Spectrum of Two-Dimensional Turbulence on a Rotating Sphere and Some Basic Features of Atmospheric Circulation on Giant Planets},
  author={Sukoriansky, Semion and Galperin, Boris and Dikovskaya, Nadejda},
  journal={Physical Review Letters},
  volume={89},
  number={12},
  pages={124501},
  year={2002},
  publisher={APS},
  doi={10.1103/PhysRevLett.89.124501}
}

@article{galperin2010geophysical,
  title={Geophysical flows with anisotropic turbulence and dispersive waves: flows with a $\beta$-effect},
  author={Galperin, Boris and Sukoriansky, Semion and Dikovskaya, Nadejda},
  journal={Ocean Dynamics},
  volume={60},
  pages={427--441},
  year={2010},
  publisher={Springer},
  doi={10.1007/s10236-010-0278-2}
}

@article{boffetta2012two,
  title={Two-dimensional turbulence},
  author={Boffetta, Guido and Ecke, Robert E},
  journal={Annual Review of Fluid Mechanics},
  volume={44},
  number={1},
  pages={427--451},
  year={2012},
  publisher={Annual Reviews},
  doi={10.1146/annurev-fluid-120710-101240}
}

@article{sukoriansky2007arrest,
  title={On the arrest of inverse energy cascade and the {R}hines scale},
  author={Sukoriansky, Semion and Dikovskaya, Nadejda and Galperin, Boris},
  journal={Journal of the Atmospheric Sciences},
  volume={64},
  number={9},
  pages={3312--3327},
  year={2007},
  doi={10.1175/JAS4013.1}
}

@article{galperin2014cassini,
  title={Cassini observations reveal a regime of zonostrophic macroturbulence on {J}upiter},
  author={Galperin, Boris and Young, Roland MB and Sukoriansky, Semion and Dikovskaya, Nadejda and Read, Peter L and Lancaster, Andrew J and Armstrong, David},
  journal={Icarus},
  volume={229},
  pages={295--320},
  year={2014},
  publisher={Elsevier},
  doi={10.1016/j.icarus.2013.08.030}
}

@article{cabanes2024zonostrophic,
  title={Zonostrophic turbulence in the subsurface oceans of the {J}ovian and {S}aturnian moons},
  author={Cabanes, Simon and Gastine, Thomas and Fournier, Alexandre},
  journal={Icarus},
  volume={415},
  pages={116047},
  year={2024},
  publisher={Elsevier},
  doi={10.1016/j.icarus.2024.116047}
}

@article{wordsworth2008turbulence,
  title={Turbulence, waves, and jets in a differentially heated rotating annulus experiment},
  author={Wordsworth, RD and Read, PL and Yamazaki, YH},
  journal={Physics of Fluids},
  volume={20},
  number={12},
  year={2008},
  publisher={AIP Publishing},
  doi={10.1063/1.2990042}
}

@article{macrobert1932xvi,
  title={{XVI}.—{F}ourier integrals},
  author={MacRobert, TM},
  journal={Proceedings of the Royal Society of Edinburgh},
  volume={51},
  pages={116--126},
  year={1932},
  publisher={Royal Society of Edinburgh Scotland Foundation},
  doi={10.1017/S0370164600023063}
}

@article{cinelli1965extension,
  title={An extension of the finite {H}ankel transform and applications},
  author={Cinelli, G1},
  journal={International Journal of Engineering Science},
  volume={3},
  number={5},
  pages={539--559},
  year={1965},
  publisher={Elsevier},
  doi={10.1016/0020-7225(65)90034-0}
}

@article{boffetta2007energy,
  title={Energy and enstrophy fluxes in the double cascade of two-dimensional turbulence},
  author={Boffetta, Guido},
  journal={Journal of Fluid Mechanics},
  volume={589},
  pages={253--260},
  year={2007},
  publisher={Cambridge University Press},
  doi={10.1017/S0022112007008014}
}

@article{lemasquerier2021zonal,
  title={Zonal jets at the laboratory scale: hysteresis and Rossby waves resonance},
  author={Lemasquerier, Daphn{\'e} and Favier, Benjamin and Le Bars, Michael},
  journal={Journal of Fluid Mechanics},
  volume={910},
  pages={A18},
  year={2021},
  publisher={Cambridge University Press},
  doi={10.1017/jfm.2020.1000}
}

@article{simonnet2021multistability,
  title={Multistability and rare spontaneous transitions in barotropic $\beta$-plane turbulence},
  author={Simonnet, Eric and Rolland, Joran and Bouchet, Freddy},
  journal={Journal of the Atmospheric Sciences},
  volume={78},
  number={6},
  pages={1889--1911},
  year={2021},
  doi={10.1175/JAS-D-20-0279.1}
}

@article{shokar2024stochastic,
  title={Stochastic latent transformer: Efficient modeling of stochastically forced zonal jets},
  author={Shokar, Ira JS and Kerswell, Rich R and Haynes, Peter H},
  journal={Journal of Advances in Modeling Earth Systems},
  volume={16},
  number={6},
  pages={e2023MS004177},
  year={2024},
  publisher={Wiley Online Library},
  doi={10.1029/2023MS004177}
}

@article{marti2016computationally,
  title={A computationally efficient spectral method for modeling core dynamics},
  author={Marti, P and Calkins, MA and Julien, K},
  journal={Geochemistry, Geophysics, Geosystems},
  volume={17},
  number={8},
  pages={3031--3053},
  year={2016},
  publisher={Wiley Online Library},
  doi={10.1002/2016GC006438}
}

@book{boyd2001chebyshev,
  title={Chebyshev and {F}ourier {S}pectral {M}ethods},
  author={Boyd, John P},
  edition={2},
  year={2001},
  publisher={Dover},
  address={Mineola}
}

@book{canuto2006spectral,
  title={Spectral {M}ethods: {F}undamentals in {S}ingle {D}omains},
  author={Canuto, Claudio and Hussaini, M Youssuff and Quarteroni, Alfio and Zang, Thomas A},
  volume={285},
  series={Scientific Computation},
  year={2006},
  publisher={Springer},
  address={Berlin, Heidelberg},
  doi={10.1007/978-3-540-30726-6}
}

@article{glatzmaier1984numerical,
  title={Numerical simulations of stellar convective dynamos. {I}. {T}he model and method},
  author={Glatzmaier, Gary A},
  journal={Journal of Computational Physics},
  volume={55},
  number={3},
  pages={461--484},
  year={1984},
  publisher={Elsevier},
  doi={10.1016/0021-9991(84)90033-0}
}

@software{hugo_frezat_2026_22915270,
  author={Hugo Frezat},
  title={hrkz/online-sgs-qg-planetary},
  month=nov,
  year=2026,
  publisher={Zenodo},
  doi={10.5281/zenodo.22915270},
  url={https://doi.org/10.5281/zenodo.22915270},
}

\end{document}